\documentclass[pre,twocolumn,showpacs,groupaddress,preprintnumbers,floatfix]{revtex4-1}

\usepackage[margin=1in]{geometry}
\usepackage{etex}
\usepackage{ifpdf}
\usepackage{calc}
\usepackage{xifthen}
\usepackage{xkeyval}
\usepackage{mathtools}
\usepackage[pagebackref]{hyperref}
\usepackage{dcolumn}
\usepackage{url}
\usepackage{mathtools}\usepackage{amsmath}
\usepackage{amsfonts}
\usepackage{amssymb}
\usepackage{amsthm}
\usepackage{array}
\usepackage{bm}   
\usepackage{bbm}
\usepackage{verbatim}
\usepackage{stmaryrd}
\usepackage{xcolor}
\usepackage{setspace}
\usepackage{graphicx}

\def\baseimagedir{img/}
\def\tikzpath{\baseimagedir tikz/}
\newcommand{\tikzpic}[2][]{\includegraphics[#1]{\tikzpath #2}}

\newcommand{\mupi} {\mu_{\pi}}

\theoremstyle{plain}	\newtheorem{Lem}{Lemma}
\theoremstyle{plain}	\newtheorem*{ProLem}{Proof}
\theoremstyle{plain} 	
\theoremstyle{plain} 	
\theoremstyle{plain} 	\newtheorem{The}{Theorem}
\theoremstyle{plain} 	
\theoremstyle{plain} 	
\theoremstyle{plain} 	
\theoremstyle{plain} 	
\theoremstyle{plain}	
\theoremstyle{plain}	\newtheorem{Def}{Definition}
\theoremstyle{plain}	

\def\clap#1{\hbox to 0pt{\hss#1\hss}}

\def\mathrlap{\mathpalette\mathrlapinternal}

\def\mathrlapinternal#1#2{%
\rlap{$\mathsurround=0pt#1{#2}$}}

\newcommand{\eM}     {\mbox{$\epsilon$-machine}}
\newcommand{\eMs}    {\mbox{$\epsilon$-machines}}
\newcommand{\EM}     {\mbox{$\epsilon$-Machine}}

\newcommand{\Process}{\mathcal{P}}

\newcommand{\MeasAlphabet}	{\mathcal{A}}
\newcommand{\MeasSymbol}   { {X} }
\newcommand{\meassymbol}   { {x} }

\newcommand{\Past}	{ \smash{\overleftarrow {\MeasSymbol}} }
\newcommand{\past}	{ \smash{\overleftarrow {\meassymbol}} }

\newcommand{\Future}	{ \smash{\overrightarrow{\MeasSymbol}} }
\newcommand{\future}	{ \smash{\overrightarrow{\meassymbol}} }

\newcommand{\AllPasts}	    { \smash{\overleftarrow{ {\rm {\bf \MeasSymbol}} } } }

\newcommand{\CausalState}	{ \mathcal{S} }

\newcommand{\causalstate}	{ \sigma }
\newcommand{\CausalStateSet}	{ \boldsymbol{\CausalState} }
\newcommand{\AlternateState}	{ R }

\newcommand{\PrescientState}	{ \widehat{\AlternateState} }

\newcommand{\PrescientStateSet}	{ \boldsymbol{\PrescientState}}

\newcommand{\Prob}      {\Pr} 

\newcommand{\Cmu}		{C_\mu}
\newcommand{\hmu}		{h_\mu}
\newcommand{\EE}		{{\bf E}}
\newcommand{\TI}		{{\bf T}}

\newcommand{\PC}		{\chi}

\newcommand{\ProcessAlphabet}	{\MeasAlphabet}

\newcommand{\forward}{+}
\newcommand{\reverse}{-}
\newcommand{\forwardreverse}{\pm} 

\newcommand{\FutureCausalState}	{ {\CausalState}^{\forward} }

\newcommand{\PastCausalState}	{ {\CausalState}^{\reverse} }

\newcommand{\MSP}{\mathcal{U}}
\newcommand{\one}{\mathbf{1}}

\newcommand{\lastindex}[2]{
  \edef\tempa{0}
  \edef\tempb{#2}
  \ifx\tempa\tempb
    \edef\tempc{#1}
  \else
    \edef\tempa{0}
    \edef\tempb{#1}
    \ifx\tempa\tempb
      \edef\tempc{#2}
    \else
      \edef\tempc{#1+#2}
    \fi
  \fi
  \tempc
}

\newcommand{\CSjoint}[1][,]{
   \edef\tempa{:}
   \edef\tempb{#1}
   \ifx\tempa\tempb
      \ensuremath{\FutureCausalState\!#1\PastCausalState}
   \else
      \ensuremath{\FutureCausalState#1\PastCausalState}
   \fi
}

\newif\ifpm 
\edef\tempa{\forwardreverse}
\edef\tempb{\pm}
\ifx\tempa\tempb
   \pmfalse
\else
   \pmtrue  
\fi

\renewcommand{\Pr}{\mathbb{P}}

\MHInternalSyntaxOn
\providecommand*\phantomword[3][c]{%
  \mathchoice
  {\MT_phantom_word:NNnn #1\displaystyle {#2}{#3}}%
  {\MT_phantom_word:NNnn #1\textstyle {#2}{#3}}%
  {\MT_phantom_word:NNnn #1\scriptstyle {#2}{#3}}%
  {\MT_phantom_word:NNnn #1\scriptscriptstyle {#2}{#3}}%
}
\def\MT_phantom_word:NNnn #1#2#3#4{%
  \@begin@tempboxa\hbox{$\m@th#2#4$}%
    \setlength\@tempdima{\widthof{$\m@th#2#3$}}%
    \hbox{\hb@xt@\@tempdima{\csname bm@#1\endcsname}}%
  \@end@tempboxa}
\MHInternalSyntaxOff

\makeatletter
\newlength\CMlength
\define@cmdkey[CM]{vector}[CM@@]{raise}{}
\define@cmdkey[CM]{vector}[CM@@]{pre}{}
\define@cmdkey[CM]{vector}[CM@@]{post}{}
\define@cmdkey[CM]{vector}[CM@@]{symbol}{}
\presetkeys[CM]{vector}{%
  raise=1.02,%
  pre={\scriptscriptstyle\,},%
  post={}%
}{}
\newcommand{\CMvector}[2][]{%
  \setkeys[CM]{vector}[]{#1}%
  \settoheight{\CMlength}{\ensuremath{#2}}%
  \setlength{\CMlength}{\CM@@raise\CMlength}%
  \ensuremath{%
    \mathrlap{\ensuremath{#2}}%
    \smash{\phantomword[c]%
      {\ensuremath{#2}}%
      {\raise \CMlength \hbox{\ensuremath{\CM@@pre\CM@@symbol\CM@@post}}}%
    }%
  }%
}
\makeatother

\newcommand{\leftrightharpoonup}{\mathrlap{\,\leftharpoonup}\phantomword[l]{\,\leftharpoonup}{\,\rightharpoonup}}

\newcommand{\CMlharpoon}[1]{\CMvector[symbol=\leftharpoonup,pre=\scriptscriptstyle]{#1}}
\newcommand{\CMrharpoon}[1]{\CMvector[symbol=\rightharpoonup,pre=\scriptscriptstyle]{#1}}
\newcommand{\CMlrharpoon}[1]{\CMvector[symbol=\leftrightharpoonup,pre=\scriptscriptstyle]{#1}}

\newcommand{\processlist}[3][\relax]{%
  \def\listfinish{#1}%
  \long\def\listact{#2}%
  \processfirst#3\listfinish}
\newcommand{\processfirst}[1]{#1\expandafter\processnext}
\newcommand{\processnext}[1]{%
  \ifx\listfinish#1\empty\else\listact{#1}\expandafter\processnext\fi}

\newcommand{\MI}[2][]{%
  \def\MIsep{:}
  \def\MIcsep{\,|\,}
  \ifthenelse{\isempty{#1}}%
    {I[\processlist{\MIsep}{#2}]}%
    {I[\processlist{\MIsep}{#2}\MIcsep#1]}%
}

\makeatletter
\def\BEsep{:}
\newcommand{\CMIndexedSymbol}[2]{%
  \protected\expandafter\def\csname #1\endcsname{%
    \@ifnextchar[{\csname #1@i\endcsname}{#2}
  }%
  \expandafter\def\csname #1@i\endcsname[##1]{%
    \@ifnextchar[{\csname #1@ii\endcsname[{##1}]}{#2_{##1}}
  }%
  \expandafter\def\csname #1@ii\endcsname[##1][##2]{%
    \ifthenelse{\isempty{##1}}%
      {\ifthenelse{\isempty{##2}}%
        {\CMlrharpoon{#2}{}}
        {\CMlharpoon{#2}_{##2}}
      }%
      {\ifthenelse{\isempty{##2}}%
        {\CMrharpoon{#2}_{##1}}
        {#2_{##1\BEsep##2}}
      }%
  }%
}

\newcommand{\CMSuperIndexedSymbol}[3]{%
  \protected\expandafter\def\csname #1\endcsname{%
    \@ifnextchar[{\csname #1@i\endcsname}{#2^#3}
  }%
  \expandafter\def\csname #1@i\endcsname[##1]{%
    \@ifnextchar[{\csname #1@ii\endcsname[{##1}]}{#2_{##1}^{#3}}
  }%
  \expandafter\def\csname #1@ii\endcsname[##1][##2]{%
    \ifthenelse{\isempty{##1}}%
      {\ifthenelse{\isempty{##2}}%
        {\CMlrharpoon{#2}{}^{#3}}
        {\CMlharpoon{#2}_{##2}^{#3}}
      }%
      {\ifthenelse{\isempty{##2}}%
        {\CMrharpoon{#2}_{##1}^{#3}}
        {#2_{##1\BEsep##2}^{#3}}
      }%
  }%
}

\CMIndexedSymbol{MS}{\MeasSymbol}
\CMIndexedSymbol{ms}{\meassymbol}
\CMIndexedSymbol{CS}{\CausalState}
\CMSuperIndexedSymbol{FCS}{\CausalState}{+}
\CMSuperIndexedSymbol{RCS}{\CausalState}{-}
\CMIndexedSymbol{AS}{\AlternateState}
\CMSuperIndexedSymbol{ASPrime}{\AlternateState}{\prime}

\newcommand{\Bernoulli}[1] { B \left( #1 \right) }
\newcommand{\FC} { \Bernoulli{\tfrac{1}{2}} }

\begin{document}

\title{Way More Than the Sum of Their Parts:\\
From Statistical to Structural Mixtures}

\author{James P. Crutchfield}
\email{chaos@ucdavis.edu}
\affiliation{Complexity Sciences Center and Physics and Astronomy Department\\
University of California at Davis,\\
One Shields Avenue, Davis, CA 95616}

\date{\today}

\bibliographystyle{unsrt}

\begin{abstract}
We show that mixtures comprised of multicomponent systems typically are much
more structurally complex than the sum of their parts; sometimes, infinitely
more complex. We contrast this with the more familiar notion of statistical
mixtures, demonstrating how statistical mixtures miss key aspects of
emergent hierarchical organization. This leads us to identify a new kind of
structural complexity inherent in multicomponent systems and to draw out broad
consequences for system ergodicity.

\vspace{0.1in}

\noindent {\bf Keywords}: epsilon-machine, nonergodicity, causal state,
multistationary, excess entropy, statistical complexity, hierarchy, transients

\end{abstract}

\pacs{
02.50.-r  
89.70.+c  
05.45.Tp  
02.50.Ey  
02.50.Ga  
}
\preprint{arXiv.org:2507.XXXX [cond-mat.stat-mech]}

\maketitle
\

\tableofcontents
\setstretch{1.1}

\section{Introduction}
\label{sec:introduction}

Multicomponent systems typically are much more structurally complex than the
collection of their parts; even infinitely more so. This should be contrasted
with statistical mixtures---such as arise in the Gibbs Paradox of
thermodynamics \cite[Secs. 2-3]{Jayn92a} where gases of distinct molecular
species exhibit only a modest entropy increase upon formation due to the
uncertainty in which species one has in hand. This contrast demonstrates how
the ansatz of statistical mixtures misses key aspects of hierarchical
organization. The result, as we show, is an awareness of a new kind of
structural complexity of composite systems.

The development here focuses on the theoretical core of this basic phenomenon,
arguing that it is, in fact, quite commonplace. To appreciate this, it will be
helpful to address the motivating issues upfront.

The multicomponent systems of interest are found in several different domains,
including the entropy of mixing in thermodynamics \cite{Call85a,Kond08a}, the
change point problem in statistics \cite{Lai95a}, the attractor-basin portrait
of a dynamical system \cite{Abra92a}, Smale's basic sets
\cite{Smal67a,Bowe72a}, spatially extended systems with multiple local
attractors \cite{Hans90a}, chaotic crystallography \cite{Varn02a,Varn06c},
evolutionary dynamics \cite{Crut99a}, and adaptive and learning systems with
memory.

We introduce the concept of hidden multistationary processes to capture what is
common across these domains---a system comprising multiple locally-competing
behaviors and structures. The basic idea can be appreciated within an
experimental paradigm: multistationarity models repeated experimental trials in
which different initial conditions lead to statistically distinct behaviors.

In short, one goal is to provide a model that captures what is common among
these domains, while providing an architectural, high-level view of the
state-space organization of behaviors. In particular, we would like to analyze
how unpredictable and how structurally complex hidden multistationary processes
are, when given their components, whose unpredictability and complexity we
know.  Another goal is that the approach be constructive, allowing one to
quantitatively determine essential properties and to determine precisely what
gives rise to the emergent global complexity.

The development proceeds as follows. It first reviews statistical mixtures,
briefly recalling stochastic processes, information theory, structural
complexity, and mixed state processes. It then introduces the theory and
construction of hidden multistationary processes. This includes a canonical
minimal representation of hidden multistationary processes and a method to
analyze their ergodic decompositions that determines how the latter affect
information measures.

The sections following this explore a number of examples going from the
simplest cases and familiar structured stationary component processes to the
\emph{Mother of All Processes} that subsumes them all. Taken altogether, these
illustrate a new kind of structural hierarchy and make plain how infinite
complexity naturally emerges. The development concludes drawing out parallels
with related results and consequences in nonequilibrium thermodynamics and
machine learning.

\section{Background}
\label{sec:Background}

To get started, we give a minimal summary of the required background---a
summary that assumes familiarity with computational mechanics
\cite{Crut08a,Crut08b} and with information theory for complex
systems \cite{Cove06a,Crut01a}.

\subsection{Processes}
\label{sec:Processes}

A process, denoted $\Process$, is specified by the joint distribution
$\Prob(\Past_t,\Future_t)$ over its chain of random variables $\ldots
\MeasSymbol_{-1} \MeasSymbol_{0} \MeasSymbol_{1} \ldots$. We view $\Process$ as
a \emph{communication channel} with a fixed input distribution
$\Prob(\Past_t)$: It transmits information from the \emph{past} $\Past_t =
\ldots \MeasSymbol_{t-3} \MeasSymbol_{t-2} \MeasSymbol_{t-1}$ to the
\emph{future} $\Future_t = \MeasSymbol_t \MeasSymbol_{t+1} \MeasSymbol_{t+2}
\ldots$ by storing it in the present. $\MeasSymbol_t$ denotes the discrete
random variable at time $t$ taking on values $\meassymbol$ from a discrete
alphabet $\MeasAlphabet$. And, $\MeasSymbol_t^\ell = \MeasSymbol_t
\MeasSymbol_{t+1} \ldots \MeasSymbol_{t+\ell-1}$ is the block of $\ell$ random
variables starting at time $t$. A particular realization is denoted using
lowercase: $\MeasSymbol_t^\ell = \meassymbol_t^\ell \in \MeasAlphabet^\ell$.
Often, we simply refer to a particular sequence $w = \meassymbol_0
\meassymbol_1 \ldots \meassymbol_{\ell-1}$, $x_i \in \MeasAlphabet$, as a
\emph{word}. If we have a symbol $\meassymbol$ and a word $w$, we form a new
word by concatenation: e.g., $wx$ or $xw$.

\subsection{Information}
\label{sec:Info}

Given a process, we form the block distributions $\{ \Prob(\MeasSymbol_t^\ell):
\text{for~all} ~t~\text{and}~ \ell \}$ by marginalizing the given joint
distribution:
\begin{align*}
\Prob(\MeasSymbol_t^\ell) = \sum_{\{\past_t,\future_{t+\ell}\}}
	\Prob(\past_t,\future_{t+\ell}) ~.
\end{align*}
(We ignore here the measure-theoretic construction of cylinder sets and their
measures; for background see Ref. \cite{Loom21b} and references therein.) A
stationary process is one for which $\Prob(\MeasSymbol_t^\ell) =
\Prob(\MeasSymbol_0^\ell)$ for all $t$ and $\ell$. For a stationary process, we
drop the time index and thereby have the family of \emph{word distributions}
$\Prob(\MeasSymbol^\ell)$ that completely characterizes the process.

The amount of Shannon information in words is measured by the \emph{block
entropy}:
\begin{align*}
H(\ell) = H[\Prob(\MeasSymbol^\ell)] ~,
\end{align*}
where $H[\Pr(Y)] = - \sum_{\{y\}} \Prob(y) \log_2 \Prob(y)$ is the Shannon
entropy of the random variable $Y$. A process' information production is given
by its \emph{entropy rate}:
\begin{align*}
\hmu = \lim_{\ell \to \infty} \frac{H(\ell)}{\ell} ~.
\end{align*}
It is often used to measure a process' degree of unpredictability.

At a minimum, a good predictor---denote its random variables $\PrescientState$---must capture \emph{all} of a process'
\emph{excess entropy} $\EE$ ~\cite{Crut01a}---all of the information shared
between past and future: $\EE = I[\Past;\Future]$. Here, $I[Y;Z]$ is the mutual
information between variables $Y$ and $Z$. That is, for a
good predictor $\PrescientState$: $\EE = I[\PrescientState;\Future]$.

These quantities are closely related. In particular, for finitary processes,
those with $\EE < \infty$, the block entropy has the linear asymptotic behavior:
\begin{align*}
H(\ell) \propto_{\ell \to \infty} \EE + \hmu \ell ~.
\end{align*}
More precisely:
\begin{align*}
\EE = \lim_{\ell \to \infty} \left[ H(\ell) - \hmu \ell \right] ~.
\end{align*}
This shows that $\EE$ controls the convergence of entropy rate estimates
$\hmu(\ell) = H(\ell) - H(\ell-1)$. In fact, for one-dimensional processes,
$\EE$ can also be defined in terms of entropy convergence:
\begin{align}
\EE = \sum_{\ell = 1}^\infty \left[ \hmu(\ell) - \hmu \right] ~.
\label{eq:EEviaConvergence}
\end{align}

An analogous quantity that controls the block entropy convergence to the
linear asymptote is the \emph{transient information}:
\begin{align*}
\TI = \sum_{L = 0}^{\infty} \left[ \EE + \hmu L - H(L) \right] ~.
\end{align*}
$\TI$ measures the average amount of information an observer must extract
in order to know a process' internal state. (For a review of these and related
informations see Ref. \cite{Crut01a}.)


\subsection{Structure}
\label{sec:Structure}

We refer to a model of a process---a particular choice of
$\PrescientState$---as a \emph{presentation}. Note that building a model of a
process is more demanding than developing a prediction scheme, since one wishes
to go beyond sequence statistics to express a process' mechanisms and internal
organization.

To do this, we first recall that a process' communication channel is determined
by the conditional distributions $\Prob(\Future_t|\Past_t)$. Based on this,
computational mechanics introduced an equivalence relation $\past \sim_\epsilon
\past^\prime$ that groups all of a process' histories which give rise to the
same prediction. The result is a map $\epsilon: \AllPasts \to \CausalStateSet$
from pasts to \emph{causal states} defined by:
\begin{align}
\epsilon(\past) =
  \{ \past^\prime: \Prob(\Future|\past) = \Prob(\Future|\past^\prime) \} ~.
\label{eq:CausalEquiv}
\end{align}
In other words, a process' causal states are equivalence
classes---$\CausalStateSet = \Prob(\Past,\Future) / \!\!  \sim_\epsilon$---that partition the
space $\AllPasts$ of pasts into sets which are predictively equivalent.
With the causal states in hand, one determines the causal-state to
causal-state transitions:
\begin{align*}
\{ T^{(\meassymbol)}_{\causalstate,\causalstate^\prime}:
\meassymbol \in \ProcessAlphabet,
\causalstate,\causalstate^\prime \in \CausalStateSet
  \}
  ~.
\end{align*}
The resulting model $M$, consisting of the causal states and transitions, is
called the process' \emph{\eM} \cite{CompMechMerge}:
\begin{align*}
M(\Process) \equiv \left\{ \CausalStateSet, \{T^{(\meassymbol)},
\meassymbol \in \ProcessAlphabet \} \right\}
  ~.
\end{align*}

Informally, a process is \emph{ergodic} if its statistics can be estimated from
a single realization that is sufficiently long. If $\Process$ is ergodic, then
$M(\Process)$'s recurrent causal states are strongly connected and their
asymptotic invariant distribution $\pi = \Prob(\CausalState)$ is unique and
given by $\pi = \pi T$, where $T = \sum_{\meassymbol \in \ProcessAlphabet}
T^{(\meassymbol)}$.

As described, an \eM\ is obtained from a process, but one can also simply
define an \eM\ and consider its generated process.  We will use both
notions in the following, as they are equivalent \cite{Trav11a}.
But why should one use the \eM\ presentation of a process in the
first place?

To summarize, out of all optimally predictive models $\PrescientStateSet$
resulting from a partition of the past---those such that
$\EE = I[\PrescientStateSet;\Future]$---the \eM\ captures the amount of
information that a process stores---the \emph{statistical complexity}
$\Cmu \equiv H[\Prob(\CausalState)]$. The excess entropy $\EE$---the information
explicitly observed in sequences---is only a lower bound on the information
$\Cmu$ that a process stores \cite{CompMechMerge}: $\EE \leq \Cmu$.
The difference $\PC = \Cmu - \EE$, called the \emph{crypticity}, measures
how the process hides its internal state information from an observer
\cite{Maho11a}.

A process' \eM\ is its minimal unifilar presentation. It is unique for the
process. Moreover, it allows a number of the process' complexity measures to
be directly and efficiently calculated \cite{Crut13a}. The latter include the
process' entropy rate, excess entropy, statistical complexity, and crypticity.
In short, a process' \eM\ captures all of its informational and structural
properties.


\section{Mixed State Operator}
\label{sec:MSP}

Given an \eM\ $M$, its causal states can be treated as a standard basis
$\{\mathbf{e}_j\}$ in a vector space. Then, any distribution
$\mu = \Prob(\CausalState)$ over the states is a linear combination:
$\mu = \sum_j c_j \mathbf{e}_j$. Following Ref. \cite{Crut08b},
these distributions are called \emph{mixed states}.
For an $k$ state \eM, the mixed-state space is a $k-1$-dimensional simplex
$\Delta^{k-1}$, as the distributions $\mu \in \Delta^{k-1}$ are normalized.

Consider a special subset of mixed states. Define $\mu(w)$ as the
distribution over $M$'s states induced after observing sequence $w =
\meassymbol_0 \ldots \meassymbol_{\ell-1}$, $M$
having started with state distribution $\pi$:
\begin{align}
\mupi (w)
  &\equiv \Pr(\CausalState_\ell
  	| \MeasSymbol_0^\ell=w, \CausalState_0 \sim \pi)  \nonumber \\
  & = \frac{\Pr(\MeasSymbol_0^\ell=w,
  	\CausalState_\ell,\CausalState_0 \sim \pi)}
	{\Pr(\MeasSymbol_0^\ell=w,\CausalState_0 \sim \pi)} \nonumber \\
  & = \frac{\pi T^{(w)}}{\pi T^{(w)} \one}
  ~,
\label{eq:mixedstates}
\end{align}
where $\one$ is a column vector of $1$s and $T^{(w)} =
T^{(\meassymbol_{\ell-1})} \cdots T^{(\meassymbol_{0})}$.
Here, the notation $X \sim P$ serves to indicate that random
variable $X$ is governed by distribution $P$.

The last line gives the mixed-state $\mupi(w)$ directly in terms of the initial
state distribution $\pi$ and $M$'s transition matrices. One interpretation
is that $\mupi(w)$ represents an observer's best guess as to the process'
causal-state distribution given that it saw word $w$ and knows both the
process' \eM\ and the initial distribution $\pi$.

To determine the set of mixed states allowed by a process, we simply calculate
the set of distinct $\mupi(w)$ for all words $w \in \ProcessAlphabet^*$. This
is most directly done by enumerating $w$ in lexicographic order: e.g., for a
binary alphabet successively choosing
$w \in \{ \lambda, 0, 1, 00, 01, 10, 11, \ldots \}$. Here, $\lambda$ is the
null word. As we will see, the mixed-state set can be finite or infinite.

If we consider the entire set of mixed states, then we construct a presentation
of the process by specifying the transition matrices:
\begin{align*}
\Pr(\meassymbol, \mupi(w\meassymbol) | \mupi(w))
  & \equiv \frac{\Pr(wx|\CausalState_0 \sim \pi)}
  {\Pr(w|\CausalState_0 \sim \pi)} \nonumber \\
  & = \mupi(w) T^{(x)} \one ~.
\end{align*}
Note that many words can induce the same mixed state.

It is useful to define a corresponding operator $\MSP$ that acts on a machine
$M$, returning its \emph{mixed-state presentation} $\MSP_\pi (M)$ under initial
distribution $\pi$. The examples to follow shortly illustrate how mixed states
and $\MSP_\pi (M)$ are calculated.

\section{Constructing Hidden Multistationary Processes}
\label{sec:ConstructMulti}

Recall that a hidden multistationary nonergodic process is one that evolves,
across successive realizations, to statistically distinct long-term behaviors.
We now introduce our model of this by giving a construction procedure. This, in
effect, defines what we mean by multistationary. We then develop several basic
properties and analyze in detail a series of example constructions to
illustrate them and their ergodic decompositions.

The main tool used to construct a hidden multistationary process is the
mixed-state operator $\MSP_\pi$. We show that this results in a canonical
presentation of a given set of stationary components. This is the
multistationary process' \eM.

\begin{Def}
A \emph{hidden multistationary process} (HMSP) is defined by the presentation
determined via the following procedure.
\begin{enumerate}
\item Specify an indexed family of component stationary ergodic processes
	$\{ \Process^i \}_{i \in I}$. Each is specified by its
	\eM\ presentation $M^i = M(\Process^i)$.
	The \eMs\ consist only of their recurrent states $\CausalStateSet^i$
	that, due to ergodicity, form a single, strongly connected set.
\item Specify the component's \emph{mixture distribution} $\pi$---the
	probability with which each will be visited (sampled):
\begin{align*}
	\pi^i = \Prob(M^i) ~.
\end{align*}
\item Finally, calculate the mixed-state presentation of the multistationary
	process:
\begin{align*}
M = \MSP_\pi \left( \bigotimes_{i \in I} M^i \right) ~,
\end{align*}
	where we take the tensor product of the measure semi-groups \cite{Kitc84a}
	specified by the component \eMs. In this way, $M$'s states and transitions
	are determined from the component \eMs\ and the mixture distribution $\pi$.
\end{enumerate}
\end{Def}
$M$, the result of the construction, determines the transient portion of a
nonergodic \eM. $M$'s recurrent components are essentially the same as those
($M^i$'s) of the original component stationary processes $\Process^i$.
That is to say, what is new in $M$ is the set of transient causal states.

Note that this construction is a stochastic analog of building recognizers
for multiregular formal languages \cite{McTa04a}.

\section{The Multistationary \EM}
\label{sec:MEM}

With the background and definitions set, we are ready to explore the properties
of multistationary nonergodic processes. We first establish the structural
properties of their \eM\ presentations and then their informational properties
via ergodic decompositions of various complexity measures.

Each component $M^i = \left\{ \CausalStateSet^i, \{ T_i^{(x)}, x \in
\ProcessAlphabet^i \} \right\}$, considered as generating its own process
$\Process^i$, has a stationary distribution $p^i$ over its states:
\begin{align*}
p_j^i = \Prob(\CausalState_j) ~, ~\CausalState_j \in \CausalStateSet^i ~.
\end{align*}
We will also write this as a vector over the multistationary process' recurrent
states, when we have a finite number of components:
\begin{align*}
\pi = \left[ \pi^1 \left(p_1^1 \ldots p_{j_1}^1 \right)
	\ldots
	\pi^k \left( p_1^k \ldots p_{j_k}^k \right) \right] ~,
\end{align*}
where $k = |I|$ and $j_i = |\CausalStateSet^i|$. The stationary state
distribution $\pi_{ij}$ for the multistationary process generated by $M$ is,
then:
\begin{align}
\pi_{ij} = \pi^i \cdot p_j^i ~.
\label{eq:MS_StationaryDist}
\end{align}

Consider the following properties of a multistationary process as just defined.

\begin{Lem}[Stationarity]
The state distribution $\pi_{ij}$ is stationary.
\end{Lem}

\begin{ProLem}
This follows from realizing that the recurrent portion of $M$'s transition
matrix is block diagonal. That is, asymptotically the components are
independent and, by assumption, the component distributions are invariant.
\end{ProLem}

\begin{Lem}[Unifilarity]
The hidden multistationary process machine $M$ is unifilar.
\end{Lem}

\begin{Lem}[Minimality]
The hidden multistationary process machine $M$ is minimal.
\end{Lem}


\begin{Lem}[Uniqueness]
The hidden multistationary process machine $M$ is unique.
\end{Lem}

The relevant definitions and proofs of these closely follow those given for
\eMs\ generally; see, for example, Ref. \cite{Shal98a}. We leave the proofs for
a sequel. This all noted, these remarks constitute a proof of the following
claim.

\begin{The}
The mixed state operator applied to a mixture of (finite, ergodic)
\eMs\ produces an \eM. That is, the \eM\ for the hidden multistationary process
generated by:
\begin{align*}
M = \MSP \left(\bigotimes_{i \in I} M^i \right)
\end{align*}
is an \eM.
\end{The}

{\it Remark}:
Constructing HMSPs in this way one could start with other classes of
presentation for the ergodic component processes, such as nonunifilar
presentations---i.e., generic HMMs. However, the resulting $M$ need not be an
\eM. And, as a consequence, one could not directly calculate from such an $M$
the various complexity measures nor, lacking minimality, draw structural
conclusions about its architecture. This is one reason why we choose to specify
the component processes using \eM\ presentations. Limiting the current
construction to ergodic components specified by finite-state \eM\ presentations
serves to simplify the discussion and highlight our main results.

However, lifting these various restrictions or generalizing the previous
properties to address them would be a fruitful effort giving a much broader
characterization of the complexity of multistationary processes.

So, from here on out we assume the ergodic components are \eMs\ and ask what
properties hold for the multistationary processes so constructed. We build
processes consisting of either a finite number or countably infinite number of
components.

\section{Ergodic Decompositions}
\label{sec:ErgDecomp}

Since we are given the component processes $\{\Process^i, i \in I\}$,
what can we say about the resulting multistationary process generated by $M$?
A first step develops various kinds of ergodic decomposition that attempt to
predict $M$'s properties in terms of its ergodic components' properties. The
basic question has a very long history in ergodic and information theories. The
reader is referred to the review given in Ref. \cite{Gray74a}. Our approach
here is, on the one hand, to briefly give a flavor of several ergodic
decompositions and, on the other, compensating for that lack of rigor, to
analyze in detail a number of concrete examples.

The word distribution $\Prob(\MeasSymbol^\ell)$ for
$M = \MSP \left(\prod_{i \in I} M^i \right)$ is given by:
\begin{align*}
\Prob(\MeasSymbol^\ell)  = \sum_{i \in I} \pi^i \Prob(\MeasSymbol^\ell | M^i) ~.
\end{align*}
That is, for word $w$:
\begin{align*}
\Prob(w)  = \sum_{i \in I} \pi^i \Prob_i(w) ~,
\end{align*}
where $\Prob_i(w)$ denotes the probability that $\Process^i$ generates $w$.

Quantitatively, the HMSP's block entropy is upper bounded by the component
block entropies:
\begin{align*}
H(\ell) & = H \left[ \Prob(\MeasSymbol^\ell) \right] \\
     & = H \left[ \sum_{i \in I} \pi^i \Prob(\MeasSymbol^\ell | M^i) \right] \\
     & \leq \sum_{i \in I} \pi^i H \left[ \Prob(\MeasSymbol^\ell | M^i) \right] \\
     & = \sum_{i \in I} \pi^i H^i(\ell)
	~,
\end{align*}
where the second-to-last step employs Jensen's inequality \cite{Cove06a}
and $H^i(\ell)$ is component $\Process^i$'s block entropy.

A more insightful upper bound, though, is developed by first imagining that
the sequences generated by the ergodic components do not overlap---for example,
the $\Process^i$s have disjoint alphabets $\ProcessAlphabet^i$. Then we define
an indicator function $f$ of the process and an associated random variable
$\theta$: $\theta = f(\MeasSymbol^\ell) = i$, if
$\MeasSymbol^\ell \in \ProcessAlphabet_i^\ell$. We have:
\begin{align*}
H \left[ \MeasSymbol^\ell \right]
     & = H \left[ \MeasSymbol^\ell, f(\MeasSymbol^\ell) \right] \\
     & = H [\theta] + H \left[ \MeasSymbol^\ell| \theta \right] \\
     & = H [\theta] + \sum_{i \in I} \Prob(\theta = i)
	 	H\left[ \MeasSymbol^\ell| \theta = i \right] \\
     & = H[\pi] + \sum_{i \in I} \pi^i H\left[ \MeasSymbol^\ell| M^i \right] \\
     & = H[\pi] + \sum_{i \in I} \pi^i H^i(\ell) ~.
\end{align*}
In the general setting, however, the sequences generated by distinct components
can overlap. This reduces the number of distinct positive-probability words and
so, too, the block entropy. In this way, we see that the above equality is only
an upper bound on the HMSP's block entropy:
\begin{align}
H(\ell) \leq H[\pi] + \sum_{i \in I} \pi^i H^i(\ell) ~.
\label{eq:BlockEntropyDecomp}
\end{align}
This bound highlights the contribution of the \emph{mixture entropy} $H[\pi]$.
We return to critique this notion of ergodic decomposition later on.
For now, we draw out several useful consequences of this line of reasoning,
relying on the bound Eq. \eqref{eq:BlockEntropyDecomp}. Elsewhere we explore
tighter informational bounds on decomposition.



From this, we see that an HMSP's entropy rate $\hmu$ is simply that of its
ergodic components. Assuming the mixture entropy $H[\pi]$ is finite, we have:
\begin{align*}
\hmu & = \lim_{\ell \to \infty} \frac{H(\ell)}{\ell} \\
     & \leq \lim_{\ell \to \infty} \frac{1}{\ell}
	 	\left\{ H[\pi] + \sum_{i \in I} \pi^i H^i(\ell) \right\} \\
     & = \sum_{i \in I} \pi^i \lim_{\ell \to \infty} \frac{H^i(\ell)}{\ell} \\
     & = \sum_{i \in I} \pi^i \hmu^i ~,
\end{align*}
where we have the \emph{component entropy rate} $\hmu^i = \hmu(M^i)$.
Reference \cite{Gray74a} originally established this decomposition.

What is less intuitive, though, are various complexity measures as they apply
to HMSPs. As we will see, unlike the entropy rate, which component processes
are selected and how they relate to one another play key roles. We first
consider the ergodic decomposition for excess entropy, then for the transient
information, and finally that for the statistical complexity.

The excess entropy $\EE$ also has an ergodic decomposition. In this case,
we have:
\begin{align*}
\EE & = \lim_{\ell \to \infty} \left( H(\ell) - \hmu \ell \right) \\
    & \leq \lim_{\ell \to \infty} \left( H[\pi] + \sum_{i \in I} \pi^i H^i (\ell)
		- \ell \sum_{i \in I} \pi^i \hmu^i \right) \\
    & = H[\pi] + \sum_{i \in I} \pi^i
		\left( \lim_{\ell \to \infty} \left[ H^i (\ell) - \hmu^i \ell \right] \right) \\
    & = H[\pi] + \sum_{i \in I} \pi^i \EE^i ~,
\end{align*}
where $\EE^i$ is the excess entropy for ergodic component $i$.
The excess entropy decomposition was explored in Refs. \cite{Debo07a,Debo09a}.

Combining the entropy rate and excess entropy ergodic decompositions, we see
that the block-entropy linear
asymptotes---$H^i(\ell) \propto \EE^i + \hmu^i \ell$---have their own decomposition:
\begin{align*}
\EE + \hmu \ell & \leq
  H[\pi] + \sum_{i \in I} \pi^i \EE^i + \ell \cdot \sum_{i \in I} \pi^i \hmu^i \\
  & = H[\pi] + \sum_{i \in I} \pi^i \left( \EE^i + \hmu^i \ell \right) ~.
\end{align*}

It is a simple additional step to develop the ergodic
decomposition for the transient information:
\begin{align*}
\TI & = \sum_{\ell = 0}^\infty \left[ \EE + \hmu \ell - H(\ell) \right ] \\
    & \leq \sum_{\ell = 0}^\infty \left[
	H[\pi] + \sum_{i \in I} \pi^i \left( \EE^i + \hmu^i \ell \right) \right. \\
	& ~~~\left.  + \ell \sum_{i \in I} \pi^i \hmu^i
	- H[\pi] - \sum_{i \in I} \pi^i H^i(\ell) \right] \\
    & = \sum_{\ell = 0}^\infty
	\sum_{i \in I} \pi^i \left[ \EE^i + \hmu^i \ell - H^i(\ell) \right] \\
    & = \sum_{i \in I} \pi^i \TI^i ~.
\end{align*}
Curiously, like the entropy rate decomposition, the mixture
entropy $H[\pi]$ does not play a role.


The statistical complexity also has an ergodic decomposition:
\begin{align*}
\Cmu & = - \sum_{\causalstate \in \CausalStateSet}
        \Prob(\causalstate) \log_2 \Prob(\causalstate) \\
     & = - \sum_{i \in I} \sum_{\causalstate^i \in \CausalStateSet^i}
        \Prob(\causalstate^i) \log_2 \Prob(\causalstate^i) \\
     & = - \sum_{i \in I} \sum_{j = 0}^{|\CausalStateSet^i|-1}
        \pi_{ij} \log_2 \pi_{ij} \\
     & = - \sum_{i \in I} \sum_{j = 0}^{|\CausalStateSet^i|-1}
        \pi^i p_j^i \log_2 \pi^i p_j^i \\
     & = - \sum_{i \in I}
		\pi^i \sum_{j = 0}^{|\CausalStateSet^i|-1}
        p_j^i \left( \log_2 \pi^i + \log_2 p_j^i \right) \\
     & = - \sum_{i \in I}
		\pi^i \sum_{j = 0}^{|\CausalStateSet^i|-1} p_j^i \log_2 \pi^i
        - \sum_{i \in I}
		\pi^i \sum_{j = 0}^{|\CausalStateSet^i|-1} p_j^i \log_2 p_j^i \\
     & = - \sum_{i \in I} \pi^i \log_2 \pi^i
        - \sum_{i \in I} \pi^i \Cmu^i \\
     & = H[\pi] + \sum_{i \in I} \pi^i \Cmu^i ~,
\label{eq:CmuDecomposition}
\end{align*}
where $\Cmu^i$ are the statistical complexities of the ergodic components.  The
decomposition for statistical complexity was first noted in Ref.
\cite{Lohr09a}. Note that this decomposition does not rely on assuming an
equality as in Eq. \eqref{eq:BlockEntropyDecomp}.

Finally, the multistationary crypticity $\PC$, which measures how a process
hides state information from an observer, is also unaffected by the mixture distribution:
\begin{align*}
\PC & = \Cmu - \EE \\
    & \geq H[\pi] + \sum_{i \in I} \pi^i \Cmu^i
	- \left( H[\pi] + \sum_{i \in I} \pi^i \EE^i \right) \\
    & = \sum_{i \in I} \pi^i \left( \Cmu^i - \EE^i \right) \\
    & = \sum_{i \in I} \pi^i \PC^i ~,
\end{align*}
where $\PC^i$ is the crypticity of component $M^i$.
In this, it is similar to the entropy rate and transient
information decompositions.

\section{Structural Decompositions---Beyond Statistical}
\label{sec:BeyondDecompositions}

To emphasize, what's notable in these kinds of informational decomposition is
that, for nonergodic \eMs, we have, for example:
\begin{align*}
\Cmu > \sum_{i \in I} \pi^i \Cmu^i
  ~.
\end{align*}
That is, the global structural complexity $\Cmu$ of a multistationary process
is strictly greater than that contained in its components $\{ \Cmu^i \}$. In
short, a multistationary process is \emph{at least the sum of its parts}.
Indeed, the above inequality leaves out the entropy of mixing. But this is too
facile. As we will see, multistationary processes are much, much more.

We will see below, taking a more structural perspective going beyond the
ergodic decompositions, that the transient causal state structure is key to a
process' global organization and what sequences of observations reveal. This
leads us to call into question the interpretation and use of the preceding
kinds of ergodic decomposition.

We now show that the construction procedure can be used to answer a number of
different questions about multistationary ergodic processes. Several questions are illustrated via particular examples; others via general constructions.
The series of examples is developed incrementally to highlight the
methods and particular results, as much in isolation as possible.

We first start with processes built from finite-state ergodic components that
lead to a multistationary process that is itself finite-state. Then we analyze
the case in which finite components lead to a multistationary process with an
infinite number of states. We end with examples built from an infinite number
of finitary ergodic processes. In each case, we explore the structure of the
resulting multistationary process, its complexity measures, and its ergodic
decomposition.

\subsection{Finite Hidden Multistationary Processes}

\subsubsection{A Base Case}
\label{sec:Base}

A simple but illustrative case is that of two period-$1$ component processes:
all Heads and all Tails, selected with fair probability: $\pi = (1/2,1/2)$.

The components observed separately have $\hmu^0 = \hmu^1 = 0$. But together
$H(\ell) = 1$, $\ell \geq 1$. In this way, we see that the HMSP information
$H(\ell)$ of the mixture is all mixing entropy $H(\pi)$.

\subsubsection{Period-$1$ and Period-$2$ Process}
\label{sec:P1P2}

Define the \emph{Periodic Process} $\Process \equiv \mathrm{P}(p)$ that repeats
the word $w = 0^{p-1}1$. Let's construct the simplest multistationary process
consisting of two such components:
\begin{enumerate}
      \setlength{\topsep}{-4pt}
      \setlength{\itemsep}{-4pt}
      \setlength{\parsep}{-4pt}
\item Period-$1$ Process $\mathrm{P}(1)$, which has complexity measures
$\hmu^1 = 0$ bits per symbol, $\Cmu^1 = 0$ bits, $\EE^1 = 0$ bits,
$\TI^1 = 0$ bit-symbols, and $\PC^1 = 0$ bits.
\item Period-$2$ Process $\mathrm{P}(2)$, which has complexity measures
$\hmu^1 = 0$ bits per symbol, $\Cmu^1 = 1$ bit, $\EE^1 = 1$ bit,
$\TI^1 = 1$ bit-symbol, and $\PC^1 = 0$ bits.
\end{enumerate}
The Period-$1$ component has a single recurrent state $A$ and the Period-$2$,
two recurrent states, label them $B$ and $C$. The second step is to specify the
mixture distribution $\pi$ and we take this to be uniform: $\pi =
\left(\tfrac{1}{2},\tfrac{1}{2}\right)$. That is, $\Prob(M^1) = 1/2$ and
$\Prob(M^2) = 1/2$. And, the final step is to use the mixed-state operator to
construct $M = \MSP_\pi \left(\mathrm{P}(1) \otimes \mathrm{P}(2) \right)$.
The resulting multistationary \eM\ is shown in Fig. \ref{fig:P1P2}(c).

\begin{figure}
  \centering
  \tikzpic{Individual_P1P2}
  \tikzpic{P1P2_MaxLength_3}
  \caption{The Period-$1$ and Period-$2$ Hidden Multistationary Process:
    (a) Component $\mathrm{P}(1)$, (b) Component $\mathrm{P}(2)$, and
    (c) $M = \MSP_\pi (\mathrm{P}(1) \otimes \mathrm{P}(2))$ with
    $\pi = \left(\tfrac{1}{2},\tfrac{1}{2}\right)$. Recurrent
    causal states are shown as hollow circles and transient causal
    states as small solid (black) circles. The start state
    sports a double circle. Transitions are labeled $p|x$ to
    indicate taking the transition with probability $p$ and
    emitting symbol $x \in \ProcessAlphabet$.
    }
\label{fig:P1P2}
\end{figure}

The recurrent states of the component \eMs\ show up as $M$'s recurrent states,
as claimed. What is new is the set of two transient states (solid circles). As
a generator of the multistationary process, $M$ begins in its start state
(solid circle, with circumscribing circle) and, then, follows transitions
according to the edge probabilities, emitting the corresponding symbols.

We can understand $M$'s structure by calculating its mixed states $\mu(w) =
\left(\Pr(A), \Pr(B), \Pr(C) \right)$, $w \in \ProcessAlphabet$, using Eq.
\eqref{eq:mixedstates}:
\begin{align*}
\mu(\lambda) & = \left[ \tfrac{1}{2},\tfrac{1}{4},\tfrac{1}{4} \right] ~,\\
      \mu(0) & = \left[ 0, 1, 0 \right] = B ~,\\
      \mu(1) & = \left[ \tfrac{1}{2}, 0, \tfrac{1}{2} \right] ~,\\
     \mu(00) & = \left[ 0, 0, 0 \right] = \emptyset ~,\\
     \mu(01) & = \left[ 0, 0, 1 \right] = C ~,\\
     \mu(10) & = \left[ 0, 1, 0 \right] = B ~,~\text{and}\\
     \mu(11) & = \left[ 1, 0, 0 \right] = A ~.\\
\end{align*}
In this, on the one hand, $\mu(\lambda)$ is the start state of the mixed state presentation and its distribution gives the asymptotic invariant distribution over the component recurrent states $A$, $B$, and $C$---the state probabilities before any symbols have been generated.

On the other hand, if $\meassymbol = 0$ is generated, then we immediately
know the process is in component $P(2)$, since $P(1)$ cannot produce a $0$,
and, in particular, it is in a specific state, $B$. This is reflected in the
transient mixed state $\mu(0) = \left[ 0, 1, 0 \right]$. In fact, any time a
valid $0$ is generated we know $M$ is in state $B$. This is also seen in mixed
state $\mu(10)$, in which the last symbol generated is a $0$ and we again
obtain a $\delta$-function distribution concentrated on state $B$.

Now, there are also disallowed transitions and so disallowed words. This is
shown in mixed state $\mu(00) = \left[ 0, 0, 0 \right]$ for word $w = 00$.

More interestingly, though, is the transient mixed state $\mu(1) = \left[
\tfrac{1}{2}, 0, \tfrac{1}{2} \right]$, which indicates that, having seen a $1$
we know that $M$ cannot be in state $B$. However, the best we can say is that
it is either state $A$ (the Period-$1$ component) or in state $C$ (the
Period-$2$ component) with fair probability. It is not until we see another
symbol that we are guaranteed to know with certainty in which component $M$ is.
If $w = 11$, then $\Process$ is in $A$. Since we now know the state with
certainty, we say that $w = 0$ and $w = 11$ are \emph{synchronizing words}. In
this case, they are the minimal synchronizing words.

\begin{figure}
  \centering
  \includegraphics[width=\columnwidth]{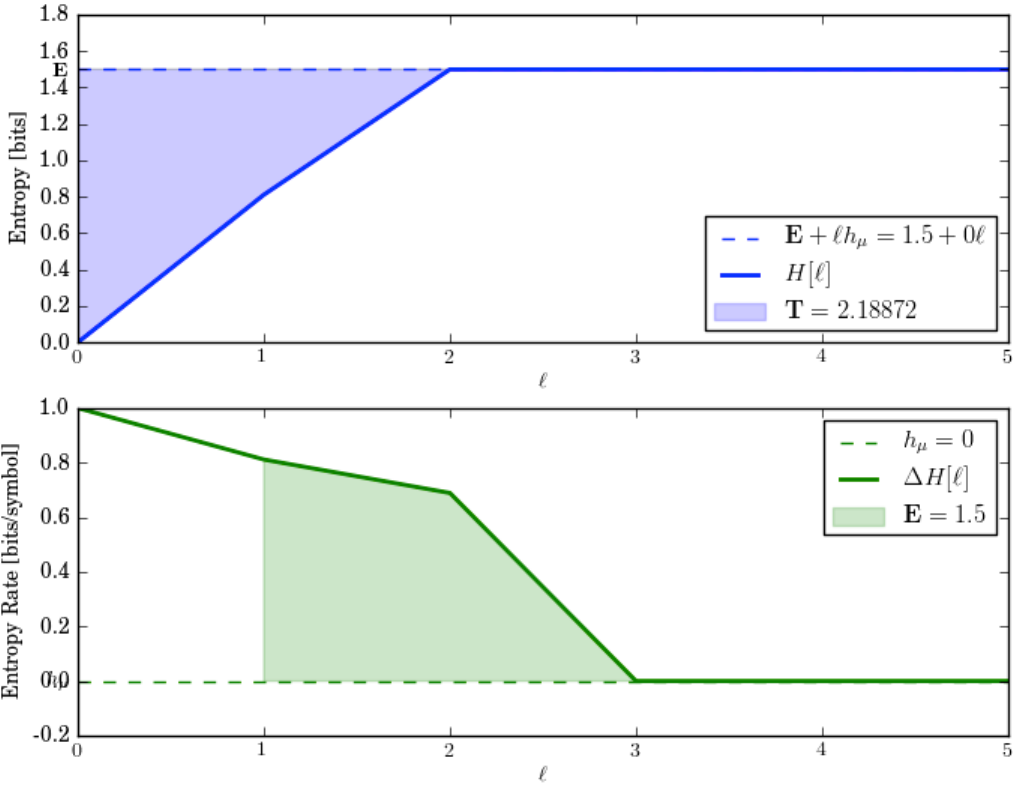}
\caption{Entropy growth $H(\ell)$ (top) and entropy convergence $\hmu(\ell)$
  (bottom) for the Period-$1$ and Period-$2$ HMSP, as function of word length
  $\ell = 0, \ldots, 5$.
  }
\label{fig:P1P2EntropyPlots}
\end{figure}

The ergodic decompositions tell us that:
\begin{enumerate}
      \setlength{\topsep}{-4pt}
      \setlength{\itemsep}{-4pt}
      \setlength{\parsep}{-4pt}
\item $\hmu = \pi^1 \hmu^1 + \pi^1 \hmu^2 = 0$ bit per symbol,
\item $\EE = H(\pi) + \pi^1 \EE^1 + \pi^2 \EE^2 = 1 + 0 + 1/2 = 1.5$ bits,
\item $\Cmu = H(\pi) + \pi^1 \Cmu^1 + \pi^2 \Cmu^2 = 1 + 0 + 1/2 = 1.5$ bits,
\item $\TI = \pi^1 \TI^1 + \pi^1 \TI^2 = 0 + 1/2 = 1/2$ bit-symbols, and
\item $\PC = \pi^1 \PC^1 + \pi^1 \PC^2 = 0 + 0 = 0$ bits.
\end{enumerate}
Let's check these by directly calculating the entropy growth $H(\ell)$ and
convergence $\hmu(\ell)$ for $M$. These are shown in Fig. \ref{fig:P1P2EntropyPlots}.

The entropy growth plot (top) leads to an estimate of $\EE \approx 1.5$ bits,
which is predicted by the ergodic decomposition. Both entropy growth and
entropy convergence (bottom) show that $\hmu(\ell) = \hmu = 0$ after $\ell =
2$. And, this too is correctly predicted by the corresponding entropy rate
decomposition.

In fact, for lengths longer than the longest period, there are always three
distinct sequences---$w \in \{1111\ldots, 0101\ldots, 1010\ldots \}$. And so,
$\EE \leq \log_2 3 \approx 1.585$ bits. This is roughly consistent with block
entropy plots.

Let's analyze this exactly.
One of those sequences is $w = 1^n$ and it occurs with probability
$1/2$. The two other sequences are $w = (01)^n$ and $w = (10)^n$ and they
are generated equally often by their component. But since that component
appears only half the time, they occur in the output sequences with probability
$1/4$ each. Thus, $\EE = H[\Prob(w)] = H[(1/2,1/4,1/4)] = 1.5$ bits. And, this
is what is seen in the plots.

The HMSP's statistical complexity is:
\begin{align}
\Cmu & = H\left[ \Prob(\CausalState) \right] \\
	 & = H\left[ (1/2,1/4,1/4) \right] \\
	 & = 3/2 \text{~bits} ~.
\end{align}
which agrees with the ergodic decomposition.

The ergodic decomposition, however, predicts $\TI = 1/2$ bit-symbols, while the
entropy growth plot shows that, in fact, $\TI \approx 2.19$ bit-symbols. So,
the ergodic decomposition for $\TI$ is incorrect. In short, we see that the
ergodic decomposition does not properly account for the state distribution's
relaxation through the transient mixed states (solid circles) in $M$; Fig.
\ref{fig:P1P2}(c). That relaxation takes longer than a single step (as the
decomposition assumes) and that increased relaxation time increases $\TI$.

Note that this is one of the simpler examples of the class of processes that
have finite transients. Let's consider one that is more complex.

\subsubsection{Isomorphic Golden Means Process}
\label{sec:GMP2}

The No-Repeated-$0$s Golden Mean Process (GMP) generates all binary sequences
except those with consecutive $0$s. When a $0$ is generated then the
probability of a $0$ or a $1$ is fair. The GMP is an order-$1$ Markov process.

\begin{figure}
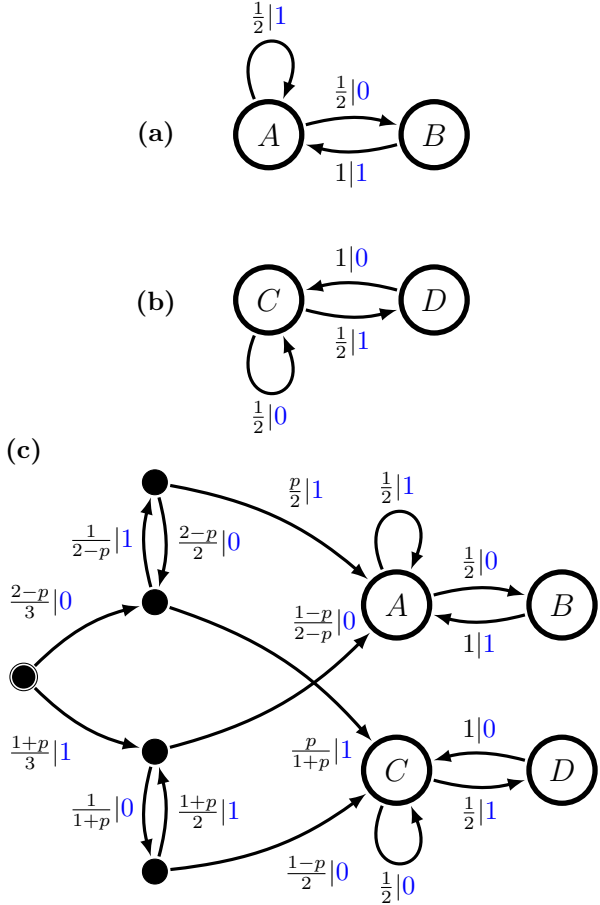

  \centering
  \tikzpic{goldenmean_recurrent}
  \tikzpic{goldenmean_full}
\caption{Two Golden Mean Processes and their nonergodic mixture:
  (a) $M^1$, (b) $M^2$, and (c) $M = \MSP_\pi (M^1 \otimes M^2)$ with
  $\pi = (p, 1-p)$.
  }
\label{fig:TwoGMPs}
\end{figure}

Let $\Process^1$ be the No-Repeated $0$s GMP, and let $\Process^2$ be the
No-Repeated-$1$s GMP. See Figs.~\ref{fig:TwoGMPs}(a) and \ref{fig:TwoGMPs}(b).
We define a nonergodic mixture $\Process$ as follows:
\begin{align*}
\Process = p \, \Process^{1} + (1-p) \, \Process^{2} ~,
\end{align*}
with mixture distribution $\pi = (p, 1-p)$.
The probability of any word $w$ is, then:
\begin{align*}\Pr(w) = p \, \Pr_1(w) + (1-p) \, \Pr_2(w) ~.
\end{align*}

Using the mixed-state operator, we construct $M(\Process)$'s transient and
recurrent states using this mixture distribution, finding:
\begin{align*}
   \mu(w) &=
          \begin{bmatrix}
            \Pr(A|w) & \Pr(B|w) & \Pr(C|w) & \Pr(D|w)
          \end{bmatrix} ~,\\
   \mu(\lambda) &=
          \begin{bmatrix}
            \frac{2p}{3} & \frac{p}{3} & \frac{2(1-p)}{3} & \frac{1-p}{3}
          \end{bmatrix} ~,\\
   \mu(0) &=
          \begin{bmatrix}
            0 & \frac{p}{2-p} & \frac{2(1-p)}{2-p} & 0
          \end{bmatrix} ~,\\
   \mu(1) &=
          \begin{bmatrix}
            \frac{2p}{1+p} & 0 & 0 & \frac{1-p}{1+p}
          \end{bmatrix} ~,\\
   \mu(00) &=
          \begin{bmatrix}
            0 & 0 & 1 & 0
          \end{bmatrix} = C~,\\
   \mu(01) &=
          \begin{bmatrix}
            p & 0 & 0 & 1-p
          \end{bmatrix}~,\\
   \mu(10) &=
          \begin{bmatrix}
            0 & p & 1-p & 0
          \end{bmatrix}~,\\
   \mu(11) &=
          \begin{bmatrix}
            1 & 0 & 0 & 0
          \end{bmatrix} = A~,\\
   \mu(001) &=
          \begin{bmatrix}
            0 & 0 & 0 & 1
          \end{bmatrix} = D~,~\text{and}\\
   \mu(110) &=
          \begin{bmatrix}
            0 & 1 & 0 & 0
          \end{bmatrix} = B ~.
\end{align*}
Longer words can only lead to one of these mixed states and so the \eM\ is
finite. The full multistationary \eM\ is shown in Fig. \ref{fig:TwoGMPs}(c),
as a function of the mixture parameter $p$.
We see that the number of states, including the transients, is
finite for all mixture probabilities.

The transition matrices for $M(\Process)$'s recurrent causal states are:
\begin{align*}
  T^0 = \begin{bmatrix}
          0 & \frac{1}{2} & 0 & 0\\
          0 & 0 & 0 & 0\\
          0 & 0 & \frac{1}{2} & 0\\
          0 & 0 & 1 & 0
         \end{bmatrix}
  \text{   and   }
  T^1 = \begin{bmatrix}
         \frac{1}{2} & 0 & 0 & 0\\
          1 & 0 & 0 & 0\\
          0 & 0 & 0 & \frac{1}{2}\\
          0 & 0 & 0 & 0
         \end{bmatrix} ~.
\end{align*}
The stationary distribution is defined by the mixture of the two processes:
\begin{align*}
\pi(p) &= \begin{pmatrix}
            p \pi^1 & (1-p) \pi^2
          \end{pmatrix}\\
       &= \tfrac{1}{3} \begin{pmatrix}
            2p & p & 2(1-p) & 1-p
          \end{pmatrix} ~,
\end{align*}
recalling that $\pi^1 = \pi^2 = \begin{pmatrix} 2/3 & 1/3 \end{pmatrix}$.

Using methods from Refs. \cite{Crut08a,Crut08b}, the excess entropy for each
recurrent component is seen to be:
\begin{align*}
\EE^1 = \EE^2 &= \frac{2}{3} \log_2 \frac{3}{2}
   	+ \frac{1}{3}\log_2{3} -     \frac{2}{3}\\
       &= \frac{2}{3} \log_2 \frac{3}{4} + \frac{1}{3}\log_2{3} \\
	   & \approx 0.251629 ~\text{bits}.
\end{align*}
By the ergodic decomposition theorem, the excess entropy for the mixture, as a  function
of $p$ is:
\begin{align*}
  \EE(p) &= p \EE^1 + (1-p) \EE^2 + H(p)\\
      &= \EE^1 + H(p) ~,
\end{align*}
since the two components are isomorphic.
For $p=1/2$, we expect $\EE \approx 1.251629$ bits.

Again, the component transient information equals the excess entropy, since the
GMP is order-$1$ Markov. So, the associated ergodic decomposition gives:
\begin{align*}
  \TI(p) &= p \TI^1 + (1-p) \TI^2 \\
      &= \TI^1 ~,
\end{align*}
since the two components are isomorphic.
For $p=1/2$, we expect $\TI \approx 0.251629$ bits.

Similarly, the statistical complexity of each recurrent component is:
\begin{align*}
  \Cmu^1 = \Cmu^2 &= \frac{2}{3} \log_2 \frac{3}{2} + \frac{1}{3}\log_2{3} \\
   & \approx 0.9182958 ~\text{bits}.
\end{align*}
So, from Eq. \eqref{eq:CmuDecomposition} the statistical complexity of the
mixture as a function of $p$ is:
\begin{align}
  \Cmu(p) = \Cmu^1 + H(p) ~.
\end{align}
For $p=1/2$, we expect $\Cmu \approx 1.9182958$ bits.

\begin{figure}
  \centering
  \includegraphics[width=\columnwidth]{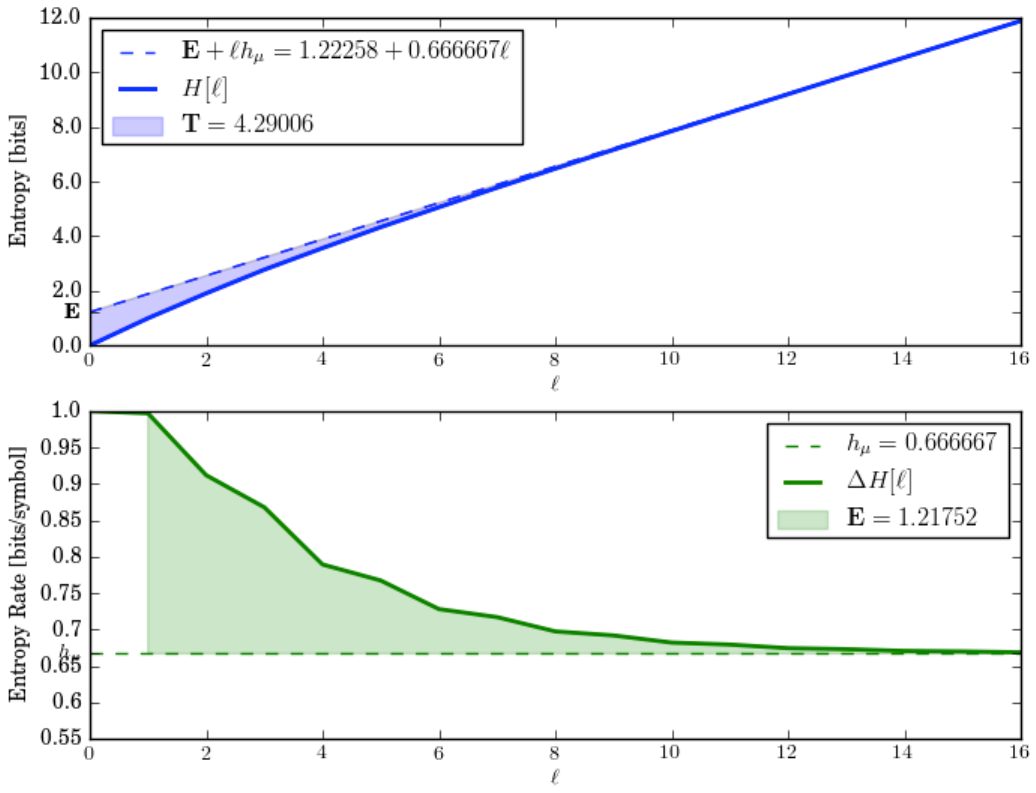}
\caption{Entropy growth $H(\ell)$ (top) and entropy convergence $\hmu(\ell)$
  (bottom) for the Two Isomorphic Golden Means HMSP, as function of word
  length $\ell = 0, \ldots, 16$ and mixture parameter $p = 1/2$.
  }
\label{fig:TwoGMPsEntropyPlots}
\end{figure}

Let's check the decompositions by calculating the associated complexity
measures from $M$'s entropy growth and convergence. The latter are shown
in Fig. \ref{fig:TwoGMPsEntropyPlots}.

The entropy growth plot estimates that $\EE = 1.22258$ bits, which is low by
2\%. And, the entropy convergence plot shows that the $\EE$, calculated there
using Eq. \eqref{eq:EEviaConvergence} as the area shown, is a bit lower still:
$\EE = 1.21753$. Although, due to the slow convergence and the finite number of
terms taken in the approximation these errors are expected. Similarly, the
ergodic decomposition of the entropy rate $\hmu = 0.811278$ bits per symbol
shows up correctly when estimated from $\Process$'s entropy growth and
convergence. And so, the predictions from the related ergodic decompositions
are consistent.

The entropy growth, however, shows the transient information is substantially
larger ($\TI \approx 4.29$ bit-symbols) than that predicted from its ergodic
decomposition ($\TI = 0.251629$). This discrepancy is clearly not due to
estimation errors. Rather, as noted above for the P1-P2 mixture, it arises from
the decomposition not accounting for the five transient causal states of $M$;
see Fig. \ref{fig:TwoGMPs}(c).

Individually, GMPs are subshifts of finite type; finite Markov order. From the
cycles in the transient states we see that as components they make the
multistationary process sofic---infinite Markov order. There are subsets of
sequences---specifically $(01)^n$---for which one never synchronizes.

This means that mixtures of finite-order Markov chains, even ``linear''
mixtures that come from independently running them, are processes that are
not finite Markovian. They require hidden Markov representations.

The preceding examples, chosen to explicitly illustrate methods and as
harbingers of coming results, are rather special in that they lead to
finite-state multistationary processes. We now turn to more typical
cases, still constructed from finite-state ergodic components, that lead
to multistationary processes with infinite states.

\subsection{Infinite State}
\label{sec:InfMultiProcess}

The preceding examples, chosen to explicitly illustrate methods and as
harbingers of coming results, are rather special in that the led to
finite-state multistationary processes. We now turn to more typical
cases, still constructed from finite-state ergodic components, that lead
to a multistationary process with an infinite number of states.

\subsubsection{Period-$1$ and Fair Coin Process}
\label{sec:FCP2}

The next example multistationary process mixes stochastic and periodic
behaviors: We build it out of a period-$1$ process and a fair coin. In effect,
we ask how difficult it is to distinguish these two simple, but extreme
processes---one completely predictable, the other completely unpredictable.

For here and a bit later, define \emph{Bernoulli Process} $\Bernoulli{p}$ which
is a model of a coin flip with bias probability $p$.

\begin{figure}
  \centering
  \tikzpic{Individual_FC_Period1}
  \tikzpic[width=\columnwidth]{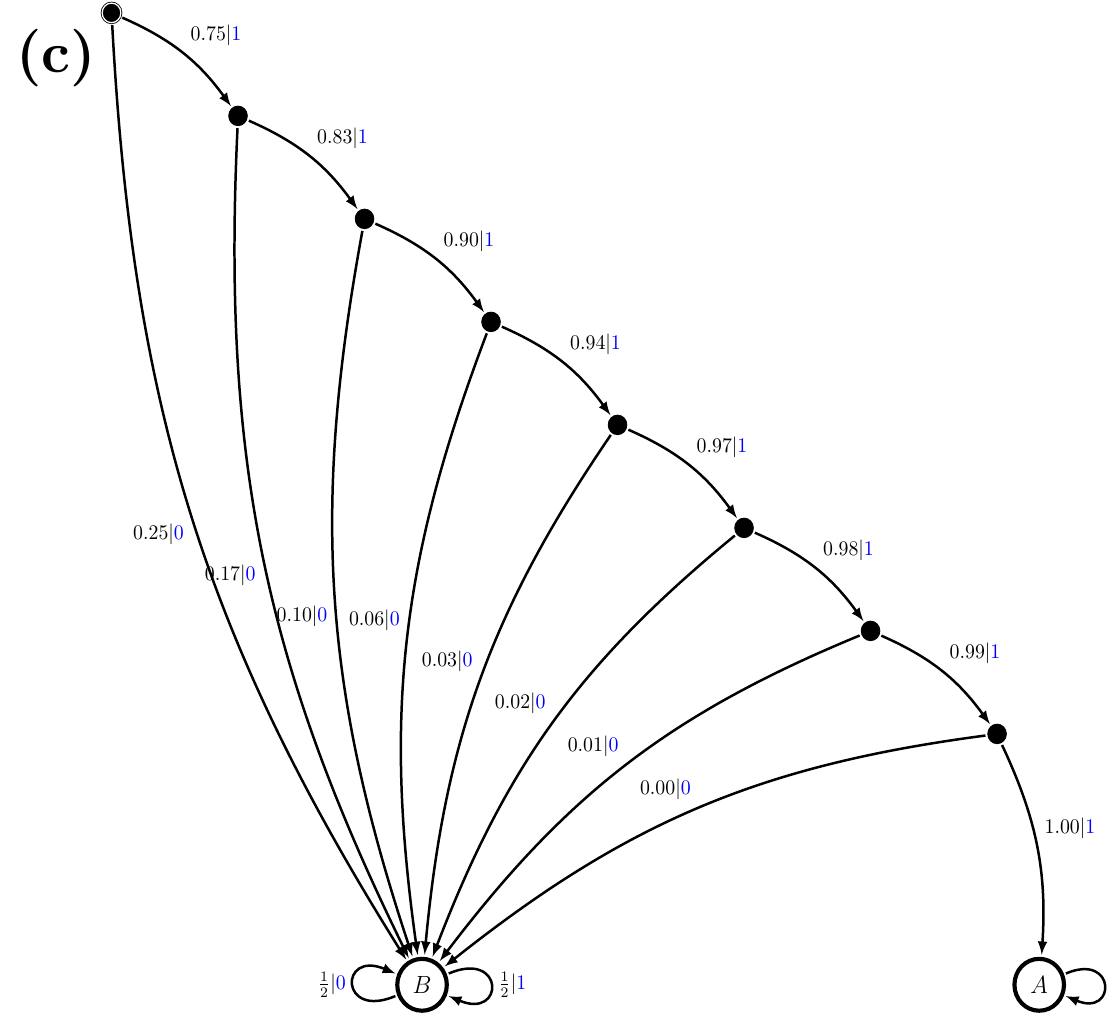}
  \caption{Period-$1$ and Fair Coin HMSP:
	(a) $\mathrm{P}(1)$, (b) $\FC$, and
	(c) $M = \MSP_\pi \left( \mathrm{P}(1) \otimes \FC \right)$ with
	$\pi = \left(\tfrac{1}{2},\tfrac{1}{2}\right)$.
	The latter is approximated by connecting the Period-$1$
	component after $8$ $1$s. In fact, the $P(1)$ component is never
	reached after any finite sequence. In contrast, $\FC$ can be reached
	quickly and by many sequences.
	Recurrent causal states are shown as hollow circles and transient causal
	states as small solid (black) circles. The start state
	sports a double circle. Transitions are labeled $p|x$ to
	indicate taking the transition with probability $p$ and
	emitting symbol $x \in \ProcessAlphabet$.
  	}
  \label{fig:FCP1}
\end{figure}

The first step, then, is to select the two stationary components:
\begin{enumerate}
      \setlength{\topsep}{-4pt}
      \setlength{\itemsep}{-4pt}
      \setlength{\parsep}{-4pt}
\item Period-$1$ Process $\mathrm{P}(1)$: $\hmu^1 = 0$ bits per symbol,
	$\Cmu^1 = 0$ bits, $\EE^1 = 0$ bits, $\TI^1 = 0$ bit-symbols,
	and $\PC^1 = 0$ bits. See Fig. \ref{fig:FCP1}(a).
\item Fair Coin Process $\FC$:, $\hmu^1 = 1$ bit per symbol, $\Cmu^1 = 0$ bits,
	$\EE^1 = 0$ bits, $\TI^1 = 0$ bit-symbols, and $\PC^1 = 0$ bits.
	See Fig. \ref{fig:FCP1}(b).
\end{enumerate}
Though at the two extremes of predictability, these are
structurally trivial processes---$\Cmu^i = 0$.

The second step is to select the mixture distribution, which we take to be
uniform: $\pi = \left(\tfrac{1}{2},\tfrac{1}{2}\right)$. And the third step is
to use the mixed-state operator to construct $M = \MSP_\pi \left(\mathrm{P}(1)
\otimes \FC \right)$. Several of the mixed states are:
\begin{align*}
\mu(\lambda) & = \left[ \tfrac{1}{2},\tfrac{1}{2} \right] ~, \\
      \mu(0) & = \left[ 0, 1 \right] = B \\
      \mu(1) & = \left[ \tfrac{3}{4}, \tfrac{1}{4} \right] ~, \\
      \mu(0\ProcessAlphabet^+) & = \left[ 0, 1 \right] = B ~, \\
      \mu(1^+0) & = \left[ 0, 1 \right] = B ~, \\
      \mu(1^k) & = \left[ 1-\alpha_k, \alpha_k \right] ~, \\
	 & \vdots ~, ~\text{and} \\
     \mu(1^\infty) & = \left[ 1, 0 \right] = A, \\
\end{align*}
where $\alpha_k = (2k+1)/2^{k+1}$.
The resulting \eM\ is shown in Fig. \ref{fig:FCP1}(c).

In Fig. \ref{fig:FCP1}(c), and so too in the mixed states, we see our first
surprising result for multistationary processes. Starting from two structurally
trivial processes, the multistationary \eM\ has a countable infinity of
transient causal states. Why? If, at any point, one sees a $0$, then we know
the process is in the Fair Coin component, since the other component cannot
generate a $0$.
However, it is only after ``seeing'' an infinite sequence of $1$s that one could
determine that the process is in the All-$1$s component. In short, the effort
required to distinguish between these two trivial processes is
infinite and this is directly reflected in the infinite set of
transient states.

\begin{figure}
  \centering
  \includegraphics[width=\columnwidth]{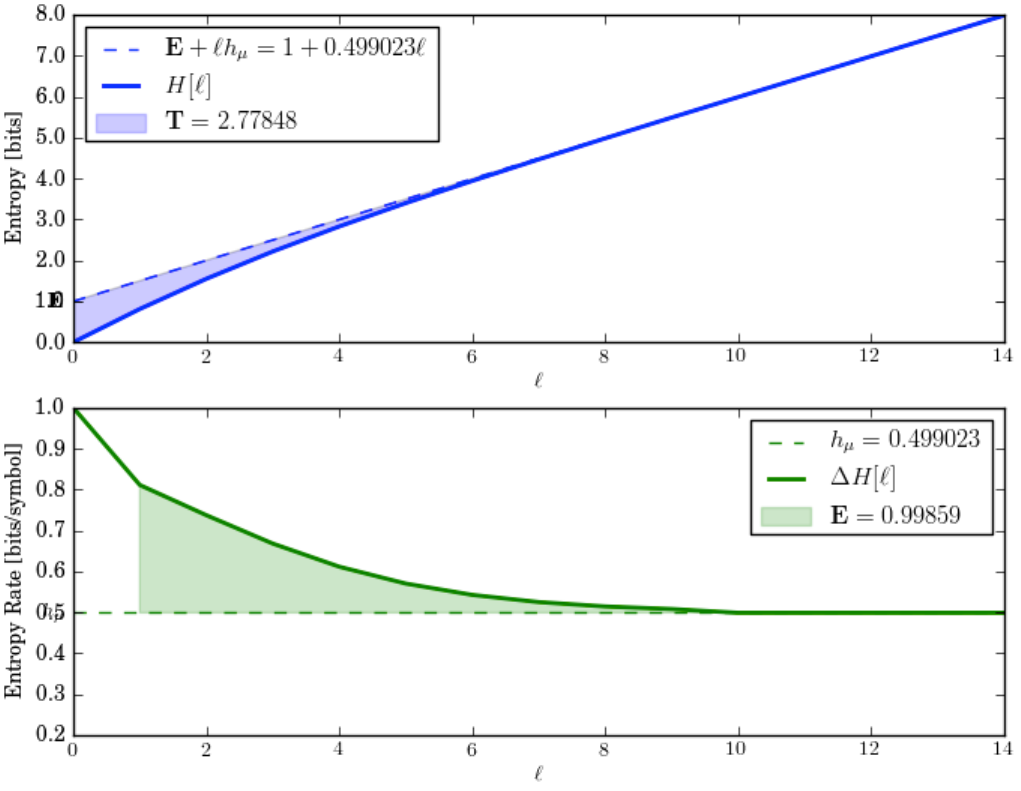}
\caption{Entropy growth $H(\ell)$ (top) and entropy convergence $\hmu(\ell)$
  (bottom) for the Period-$1$ and Fair Coin HMSP, as function of word length
  $\ell = 0, \ldots, 14$.
  }
\label{fig:FC_Period1EntropyPlots}
\end{figure}

The ergodic decompositions tell us that:
\begin{enumerate}
      \setlength{\topsep}{-4pt}
      \setlength{\itemsep}{-4pt}
      \setlength{\parsep}{-4pt}
\item $\hmu = \pi^1 \hmu^1 + \pi^2 \hmu^2 = 0 + 1/2 = 1/2$ bit per symbol,
\item $\EE = H(\pi) + \pi^1 \EE^1 + \pi^2 \EE^2 = 1 + 0 + 0 = 1$ bits,
\item $\Cmu = H(\pi) + \pi^1 \Cmu^1 + \pi^2 \Cmu^2 = 1 + 0 + 0 = 1$ bits,
\item $\TI = \pi^1 \TI^1 + \pi^2 \TI^2 = 0 + 0 = 0$ bit-symbols, and
\item $\PC = \pi^1 \PC^1 + \pi^2 \PC^2 = 0 + 0 = 0$ bits.
\end{enumerate}
Note that the ergodic decompositions predict that the structural complexity
measures are driven solely by the mixture entropy $H(\pi)$. Both components
contribute nothing: $\EE^i = \Cmu^i = 0$. Let's check these predictions by
estimating the quantities from $M$'s entropy growth and convergence, shown in
Fig. \ref{fig:FC_Period1EntropyPlots}.

The entropy growth plot shows that $\EE = 1$ bit, as predicted by $M$'s ergodic
decomposition. And, the entropy convergence plot shows that the $\EE$,
calculated as the area shown, is also the same. Similarly, the ergodic
decomposition of the entropy rate $\hmu = \tfrac{1}{2}$ bits per symbol shows
up correctly on the entropy plots.

The entropy growth plot, however, shows the transient information is quite a
bit larger ($\TI \approx 2.78$ bit-symbols) than that predicted ($\TI = 0$)
from its ergodic decomposition.

Also, the informational ergodic decompositions, while indicating a role for the
mixture entropy, miss entirely the existence of an infinite number of transient
states and the attendant difficulty that confronts an observer trying to detect
in which component the process is.

\subsubsection{Two Biased Coins}
\label{sec:BCs}

Slightly increasing the level of sophistication, we now construct a multistationary process out of fully stochastic components: Two biased coins
of unequal (but symmetric) biases $\Bernoulli{\tfrac{1}{4}}$ and
$\Bernoulli{\tfrac{3}{4}}$. See Figs. \ref{fig:TwoBCs}(a) and (b).

\begin{figure}
  \centering
  \tikzpic[width=\columnwidth]{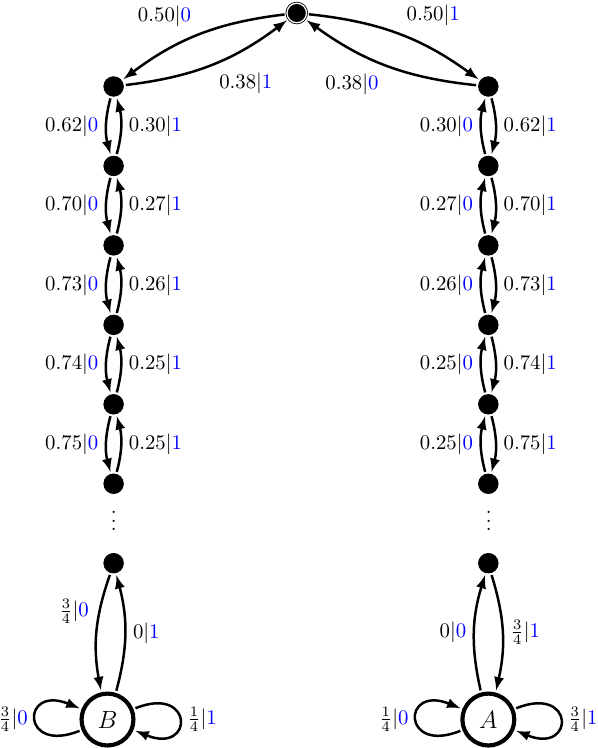}
  \caption{Two Biased Coins HMSP:
	$M = \MSP_\pi \left(\Bernoulli{\tfrac{1}{4}} \otimes
	\Bernoulli{\tfrac{3}{4}} \right)$ with
	$\pi = \left(\tfrac{1}{2},\tfrac{1}{2}\right)$.
	The latter is approximated by connecting the transient states to
	$\Bernoulli{\tfrac{1}{4}}$ and $\Bernoulli{\tfrac{3}{4}}$ after
	$30$ $0$s or $30$ $1$s, respectively. In fact,
	$\Bernoulli{\tfrac{1}{4}}$ and $\Bernoulli{\tfrac{3}{4}}$
	are never reached after any finite word.
  	}
  \label{fig:TwoBCsProcess}
\end{figure}

We again take a uniform mixture distribution:
$\pi = \left(\tfrac{1}{2},\tfrac{1}{2}\right)$.
The result of constructing $M = \MSP_\pi \left(\Bernoulli{\tfrac{1}{4}} \otimes
\Bernoulli{\tfrac{3}{4}} \right)$ with
$\pi = \left(\tfrac{1}{2},\tfrac{1}{2}\right)$
is shown in Fig. \ref{fig:TwoBCsProcess}.

\begin{figure}
  \centering
  \tikzpic{IndividualBCs}
  \caption{Two Biased Coins HMSP Components:
    (a) $\Bernoulli{\tfrac{1}{4}}$ and
    (b) $\Bernoulli{\tfrac{3}{4}}$.
    }
  \label{fig:TwoBCs}
\end{figure}

The mixed state presentation reveals two countably-infinitely-long chains of
transient causal states. One leads to the ergodic component for
$\Bernoulli{\tfrac{3}{4}}$ and the other for $\Bernoulli{\tfrac{1}{4}}$.  In a
simple sense these long transient chains show the mechanism by which one
determines the coin biases. Interestingly, though, at any point statistical
fluctuations can change the apparent bias and drive the state back up the long
chains, heading for the complementary biased coin.

Consider, as above the $\hmu$, $\EE$, $\Cmu$, $\TI$, and $\PC$ ergodic decompositions. The ergodic decompositions for excess entropy and statistical
complexity give similar results; namely:
\begin{align*}
\EE & = H[\pi] + \EE(M^1) + \EE(M^2) \\
\EE & = H[\pi] ~,~\text{and} \\
\Cmu & = H[\pi] + \Cmu(M^1) + \Cmu(M^2) \\
\Cmu & = H[\pi] ~.
\end{align*}
That is, the complexities of the multistationary process is all in the mixture
distribution. Even then, the mixture entropy, in this case a number upper
bounded by $1$, belies the infinite number of transients and the difficulty of
determining in which ergodic component the process is. Quantitatively, it seems
another measure of the global process complexity and a new decomposition are in
order. We return to this shortly, after examining several more kinds of
multistationary process.

\begin{figure}
  \centering
  \includegraphics[width=\columnwidth]{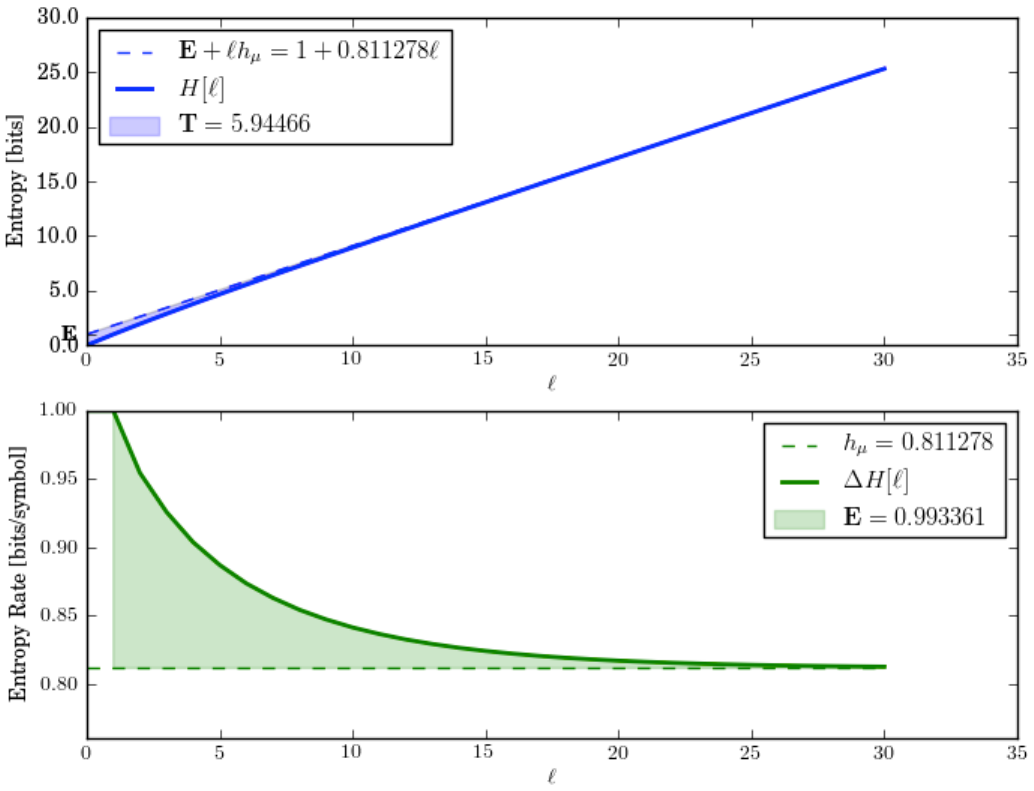}
\caption{Entropy growth $H(\ell)$ (top) and entropy convergence $\hmu(\ell)$
  (bottom) for the Two Biased Coins HMSP, as
  function of word length $\ell = 0, \ldots, 30$.
  }
\label{fig:TwoBCs_OneThreeQuartersEntropyPlots}
\end{figure}

Let's validate the ergodic decompositions' predictions vis a vis the process
estimates of their various measures from $M$'s entropy growth and convergence.
The latter are shown in Fig. \ref{fig:TwoBCs_OneThreeQuartersEntropyPlots}.

Entropy growth, using the $y$-intercept method, shows that $\EE = 1$ bit, as
predicted by $M$'s ergodic decomposition. And, the entropy convergence plot
shows that $\EE$, as the area shown, is also the same, though it takes many
terms and so shows slow convergence. The ergodic decomposition of the entropy
rate $\hmu = 0.811278$ bits per symbol shows up correctly on the entropy plots.

The entropy growth, however, shows the transient information is substantially
larger ($\TI \approx 5.95$ bit-symbols) than that predicted ($\TI = 0$)
from its ergodic decomposition. Again, the mixing entropy fails to account for
the dominating transient causal state structure.

%

\subsubsection{Pair of Isomorphic Even Processes}
\label{sec:TwoEPs}

The Even Process (EP) generates all binary sequences such that pairs of $1$s
occur in blocks of even length bounded by $0$s. Once a $0$ is seen, a $0$ or a
$1$ is generated with fair probability. The EP is closely related to the Golden
Mean Process. They have the same entropy rates and statistical complexities.
The main important difference, despite the close similarity and a simple
relabeling of transitions, is that the EP is described by no finite-order
Markov chain. It is infinite Markov order, though finite state.

To construct a multistationary process the first step, then, is to select two
stationary components. One component $\Process^1$ will be an EP with even
number of $0$s (Fig. \ref{fig:TwoEPs}(b)) and the other $\Process^2$ an EP with
an even number of $1$s (Fig. \ref{fig:TwoEPs}(a)). The second step is to choose
mixture distribution: $\pi(M^1,M^2) = (1/2,1/2)$.

Finally, Fig. \ref{fig:TwoEPMultistationary} shows the \eM\ for the HMSP $M =
\MSP_\pi \left( M^1 \otimes M^2 \right)$ with $\pi \left(M^1,M^2\right) =
\left(\tfrac{1}{2},\tfrac{1}{2}\right)$. The \eM\ displayed is estimated only up
to words of length $8$ and the transitions are set to give a well-formed \eM\
at this approximation.


There are several observations. First, the HMSP \eM\ is symmetric under $1$-$0$
exchange, as it should be given this symmetry in the ergodic components.
Second, and less obviously, there is an infinite number of transient causal
states. This is due to the outside paths along $1^\infty$ and $0^\infty$. These
two sequences arise from the $2$-cycles in the respective ergodic component
recurrent states: pairs of $1$s in $M^1$ never synchronize; ditto for pairs of
$0$s in $M^2$. And so, in $M$ there are infinitely long sequences that never
reach $M^1$ or $M^2$.

Third, the HMSP is infinite Markov order. To see this, note that there are six
cycles in the transient states---these cycles are the signature of infinite
Markov order or, what is called ``soficity''. The HMSP is a shift of infinite
type \cite{Lind95a}. In particular, there is a two-cycle $(00)^+$ between
states $32$ and $38$ and one $(11)^+$ between states $37$ and $41$. There are
two four-cycles $(1100)^+$ between states, $10$, $18$, $24$, and $26$ and
between states $24$, $29$, $34$, and $36$; and two $(0011)^+$ between states
$11$, $19$, $25$, and $23$ and between states $25$, $30$, $35$, and $33$.

\begin{figure}
  \centering
  \includegraphics[width=\columnwidth]{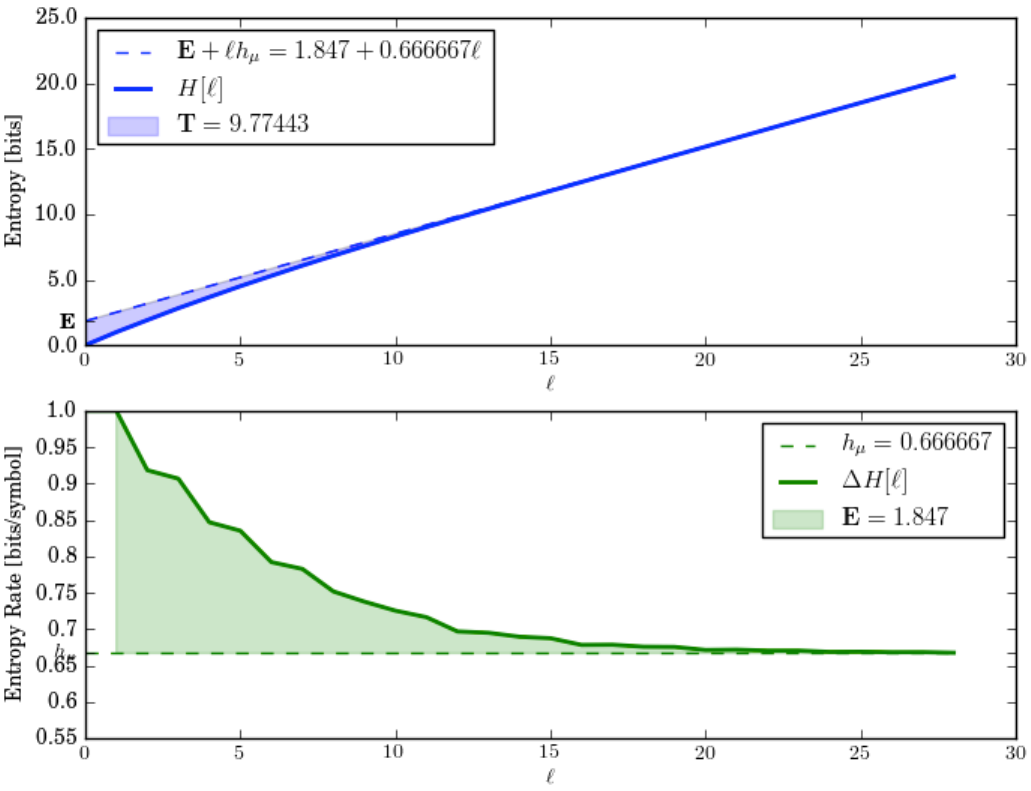}
\caption{Entropy growth $H(\ell)$ (top) and entropy convergence $\hmu(\ell)$
  (bottom) for the Two Isomorphic Even Multistationary Process, as
  function of word length $\ell = 0, \ldots, 28$.
  }
\label{fig:TwoEPsEntropyPlots}
\end{figure}

The ergodic decompositions tell us that:
\begin{enumerate}
      \setlength{\topsep}{-4pt}
      \setlength{\itemsep}{-4pt}
      \setlength{\parsep}{-4pt}
\item $\hmu = \pi^1 \hmu^1 + \pi^1 \hmu^2 = \hmu^1 = \tfrac{2}{3}$ bit per symbol,
\item $\EE = H(\pi) + \pi^1 \EE^1 + \pi^2 \EE^2 = H(\pi) + \EE^1 =
1 + 0.918296 = 1.918296$ bits,
\item $\Cmu = H(\pi) + \pi^1 \Cmu^1 + \pi^2 \Cmu^2 = H(\pi) + \Cmu^1
= 1 + 0.918296 = 1.918296$ bits,
\item $\TI = \pi^1 \TI^1 + \pi^1 \TI^2 = \TI^1 = 3.16938$ bit-symbols, and
\item $\PC = \pi^1 \PC^1 + \pi^1 \PC^2 = \PC^1 = 0 $ bits.
\end{enumerate}
Let's check these by directly calculating the entropy growth and convergence
for $M$. These are shown in Fig. \ref{fig:TwoEPsEntropyPlots}.

\begin{figure}[ht]
  \centering
  \tikzpic{TwoEPs}
\caption{Two Even Processes: (a) $M^1$, pairs of $1$s, with its two transient
  causal states and (b) $M^2$, pairs of $0$s, with its two transient causal
  states.
  }
\label{fig:TwoEPs}
\end{figure}

\begin{figure}[ht]
  \centering
  \includegraphics[width=\columnwidth]{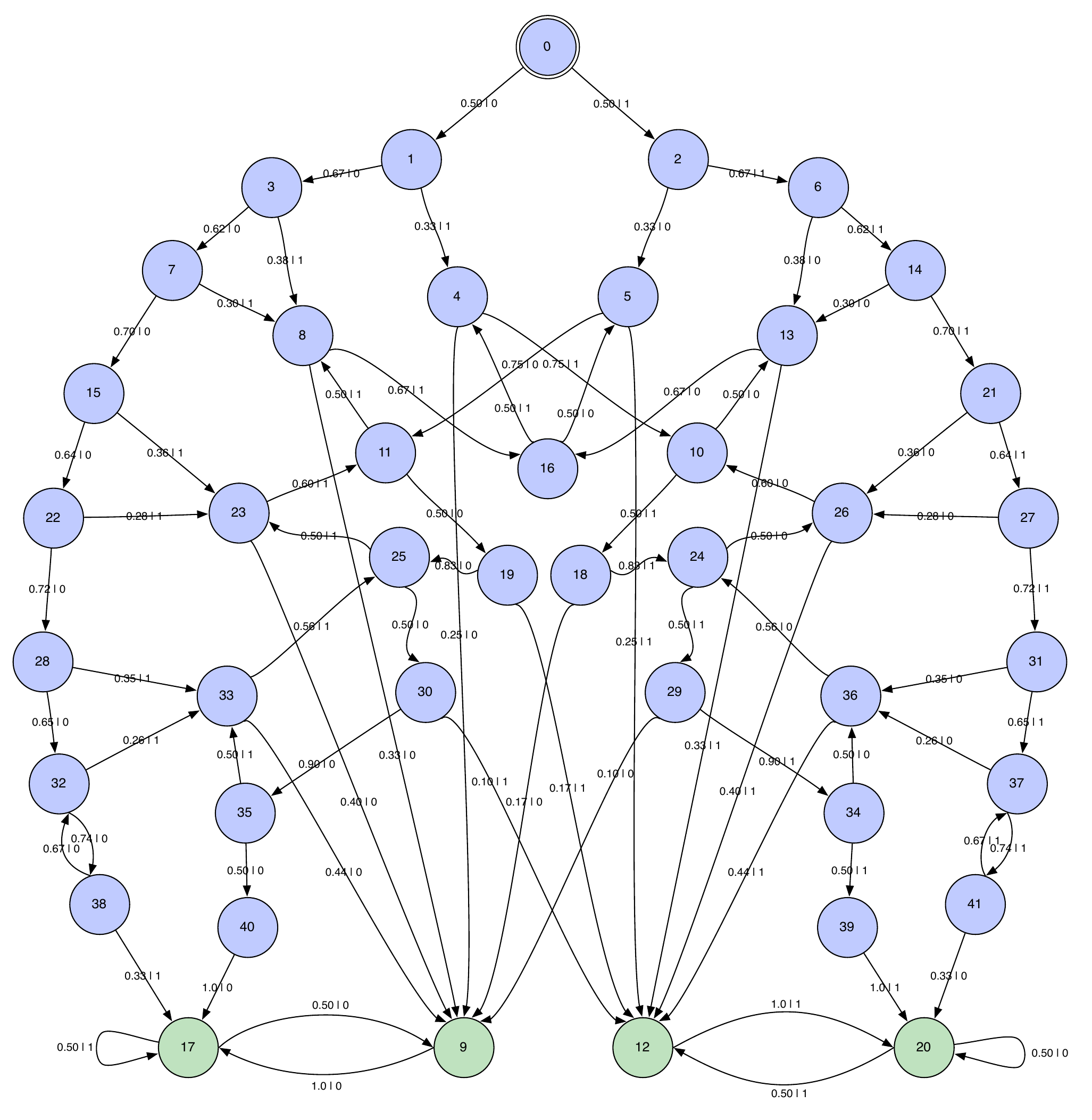}
\caption{Two Isomorphic Even Process HMSP:
  $M = \MSP_\pi \left( M^1 \otimes M^2 \right)$ with
  $\pi \left(M^1,M^2\right) = \left(\tfrac{1}{2},\tfrac{1}{2}\right)$.
  Approximated with maximum word length of $8$.
  Transient states are reconnected to mimic the component EP's transient states.
  Specifically, calculating $M$ using word length 8 leaves $M$'s states $38-41$
  as dangling states---with no outgoing transitions. Transitions were added by
  noting that these states closely approximate the EPs' two transient states,
  seen in Figs. \ref{fig:TwoEPs}(a) and (b), at sufficiently large word length.
  }
\label{fig:TwoEPMultistationary}
\end{figure}

%

Let's check the decompositions by comparing their predictions to estimates
from $M$'s entropy growth and convergence. These functions are shown in Fig.
\ref{fig:TwoEPsEntropyPlots}.

The entropy growth plot shows that $\EE = 1.847$ bits which disagrees by about
4\% with the prediction from the ergodic decomposition ($\EE = 1.918$ bits).
And, the entropy convergence plot shows that the $\EE = 1.847$ bits as the area
shown. Although, due to the slow convergence, the finite number of terms taken
in the numerical approximation, and the finite number of transient states taken
in the approximation of $M$, this error is not surprising. Similarly, the
ergodic decomposition of the entropy rate $\hmu = 2/3$ bits per symbol
shows up correctly on the entropy plots ($\hmu \approx = 0.6666$).

Entropy growth, however, shows the transient information is three times larger
($\TI \approx 9.77443$ bit-symbols) than that predicted from its ergodic
decomposition ($\TI = 3.16938$ bit-symbols). Again, this discrepancy follows
from the mixture entropy missing the contributions from the (infinite) number
of transient causal states.

\subsection{Infinite Components}
\label{sec:InfComponentProcess}

We end our selection of example multistationary processes by constructing
several from an infinite number of finitary ergodic components.

\subsubsection{Handbag of Necklaces}
\label{sec:BOP}

\begin{figure}
  \centering
  \includegraphics[width=\columnwidth]{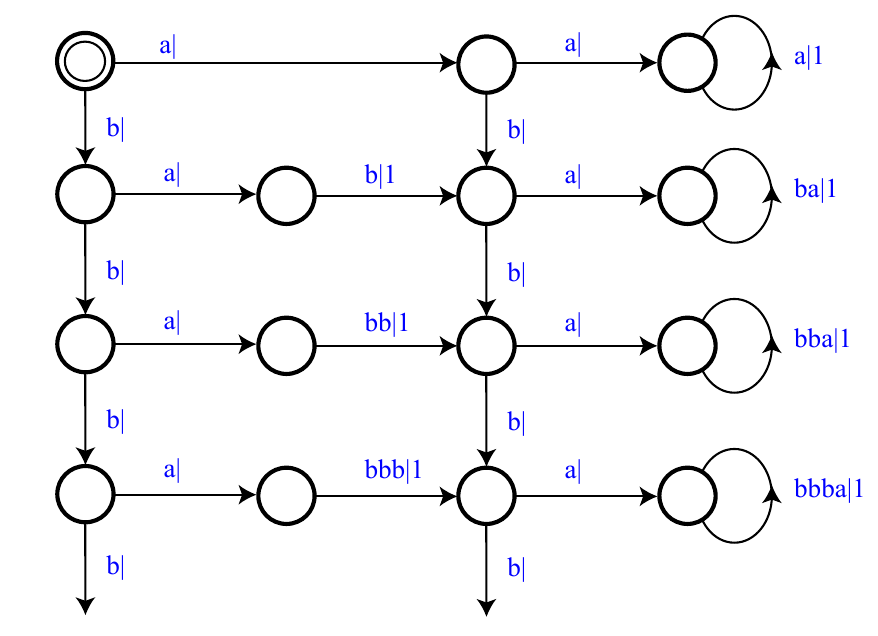}
\caption{The \eM\ for HMSP presentation of the Handbag of Necklaces Process.
  Chains of probability-$1$ transitions between causal states are given as
  strings.
  }
\label{fig:BOPeM}
\end{figure}

Fourier analysis of a signal assumes the generating process consists of at most
periodic sequences. As an analog of this assumption in the present setting,
consider the \emph{Handbag of Necklaces} (HMSP) consisting of ergodic-component
stationary processes $P(i)$ for all periods, $i \in I = 1, 2, 3, 4, \ldots$.
That is, if we assume a binary process, the sequences emitted consist of words
period-$1$ $a^+$, period-$2$ $(ab)^+$, period-$3$ $(bba)^+$, period-$4$
$(bbba)^+$, and so on. The HMSP \eM\ is shown in Fig. \ref{fig:BOPeM}.

Note that there is infinite number of transient causal states. Overall the HMSP
is a highly symmetric structure and dominated by the transient states. From
this one can readily read-off how to synchronize---how to know in which ergodic
component the process is. For example, to get to component $M^i$ there are
exactly $i$ paths.

Now, consider the mixture measure $\pi^i$ for the components. Then, the state
probabilities are $\pi_{ij} = K \pi^i / j, ~j = 1, \ldots, i$, where $K =
\sum_{i=1}^\infty \sum_{j=1}^i \pi^i / j$ is the normalization constant. Note
this is the presentations' stationary invariant distribution.

There is some flexibility in setting the mixture distribution $\pi^i$. There are
several criteria for choosing it for a countable number of states:
\begin{enumerate}
      \setlength{\topsep}{-4pt}
      \setlength{\itemsep}{-4pt}
      \setlength{\parsep}{-4pt}
\item Normalization:
	$\sum_{i=1}^\infty \sum_{j=1}^i \pi_{ij} = \sum_{i=1}^\infty \pi^i = 1$.
	And so, $\pi^i$ must decay faster than $1/i$.
\item Finitary ($\EE < \infty$) HMSP \cite{Crut01a}: Must have $H[\pi] =
	\sum_{i=1}^\infty \pi^i \log_2 \pi^i < \infty$.
\item Infinitary ($\EE \to \infty$) HMSP \cite{Crut01a}: Must have
	$H[\pi] \to \infty$. See Ref. \cite{Trav11b} for another  example process
	in this class.
\end{enumerate}

Consider the structure of the transitions in the HMSP's first row of states.
The first transition probability for seeing an $a$ is:
\begin{align*}
\Prob(a) & = \sum_{i=1}^\infty \Prob(a|M^i) \Prob(M^i) \\
         & = \sum_{i=1}^\infty \pi^i / i ~,
\end{align*}
since the probability of seeing an $a$ in the $i^{th}$ component is $1/i$.  Let
$p_a = \Prob(a)$. The probability for the succeeding transition emitting an $a$
is:
\begin{align*}
\Prob(a|a) & = \Prob(aa)/ \Prob(a) \\
           & = \pi^1 / p_a ~,
\end{align*}
since $\Prob(aa) = \pi^1$. These transitions determine those leaving the top row of states on a $b$. Note that $\Prob(a^n|aa) = 1, n = 1, 2, \ldots$.

Now, consider the second to the top row of transitions.
First, $\Prob(b) = 1-p_a$. Then, we have:
\begin{align*}
\Prob(a|b) & = \Prob(ba)/\Prob(b) \\
           & = \frac{p_a - \pi^1}{1 - p_a} ~.
\end{align*}
since:
\begin{align*}
\Prob(ba) & = \sum_{i=2}^\infty \Prob(M^i) \Prob(ba|M^i) \\
          & = \sum_{i=2}^\infty \pi^i / i \\
		  & = p_a - \pi^1 ~.
\end{align*}
Note that $\Prob(bab) = \Prob(ba)$.
There is a second path to $M^2$ controlled by the transition:
\begin{align*}
\Prob(b|a) & = \Prob(ab)/\Prob(a) \\
           & = \frac{p_a - \pi^1}{p_a} ~.
\end{align*}
since $\Prob(ab) = \Prob(ab)$. That is, the appearance of $ab$ and
$ba$ in each component occurs due to the same conditions.

The ergodic decompositions tell us that:
\begin{enumerate}
      \setlength{\topsep}{-4pt}
      \setlength{\itemsep}{-4pt}
      \setlength{\parsep}{-4pt}
\item $\hmu = \sum_{i=1}^\infty \pi^i \hmu^i = 0$, as $\hmu[P(i)] = 0$.
\item The excess-entropy ergodic decomposition for a process with $p$ component
	periodic processes with periods ${1, \ldots, p}$ is:
\begin{align}
\EE & = H[\pi] + \sum_{i=1}^p \pi^i \EE^i \\
    & = H[\pi] + \sum_{i=1}^p \pi^i \log_2 i \\
    & = \log_2 p + 1/p \sum_{i=2}^p \log_2 i ~,
\end{align}
where the second step follows assuming $\pi^i$ is uniform.
\item $\Cmu$ = \EE.
\item $\TI = \sum_{i=1}^p \pi^i \TI^i = \sum_{i=2}^p \pi^i \log_2 i$ bit-symbols, and
\item $\PC = 0$: This is a bit surprising: No crypticity, no hidden information.
\end{enumerate}

\begin{figure}
  \centering
  \includegraphics[width=\columnwidth]{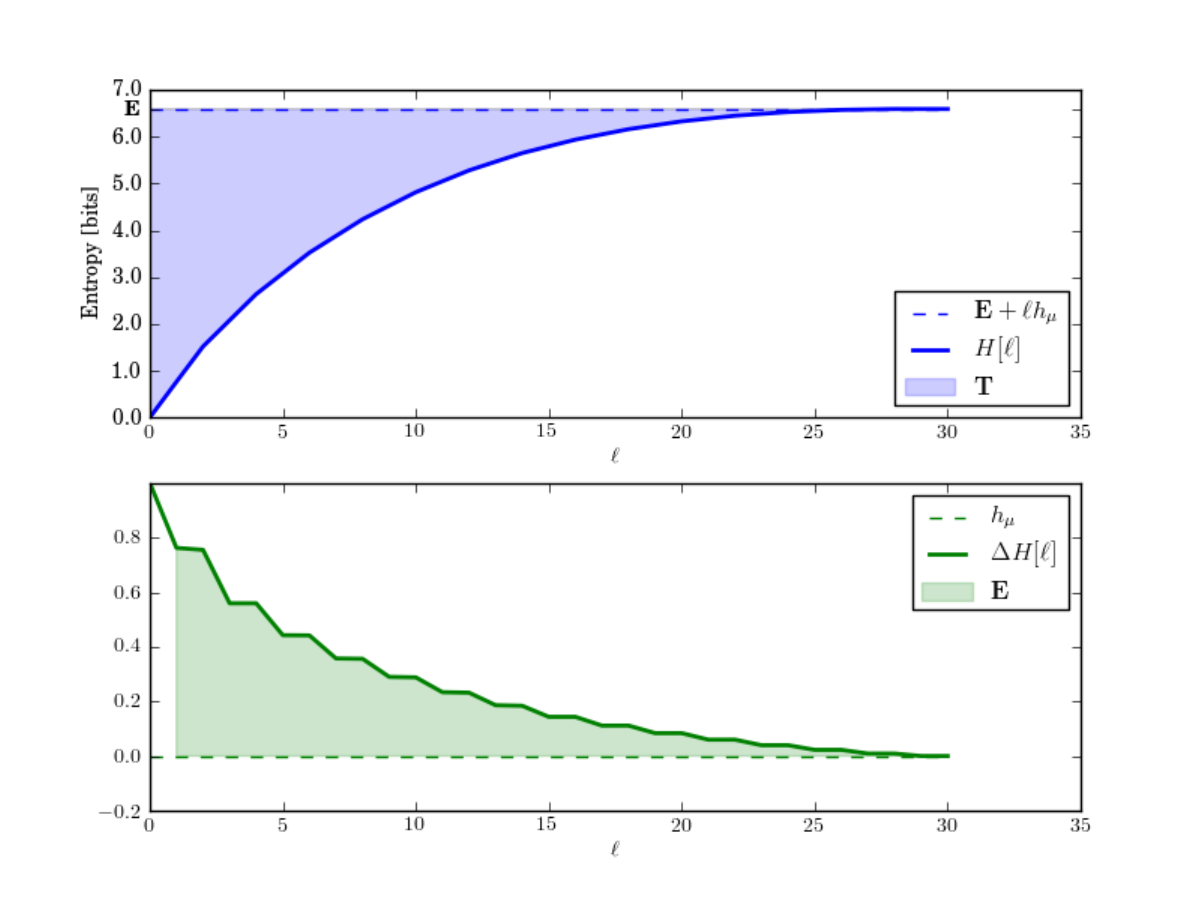}
\caption{Entropy growth $H(\ell)$ (top) and entropy convergence $\hmu(\ell)$
  (bottom) for the Handbag of Necklaces Multistationary Process, as
  function of word length $\ell = 0, \ldots, 30$ and up to period $p = 15$.
  (Note that the mixed states were approximated up to Max\_Length = $30$.)
  }
\label{fig:BOPEntropyPlots}
\end{figure}

These are consistent with directly calculating the entropy growth for
$M$, as shown in Fig. \ref{fig:BOPEntropyPlots}.


\subsubsection{The Purse Process}
\label{sec:BoC}

The example of two biased coins suggests extending to an infinite number of
biased coins in a purse---a bag of coins with different biases. As hinted at in
the two coin case, \emph{all} of the (infinite) complexity is in the mixture
and \emph{none} comes from the components.

Moreover, we can chose $\pi$ to be such that $H[\pi]$ is finite or infinite.
Thus, the Purse Process is an extreme example in which infinite complexity
comes from zero-complexity components. There is probably no simpler way to say
that a multistationary process is way more than the sum of its (zero complexity)
parts.

\begin{figure}
  \centering
  \includegraphics[width=\columnwidth]{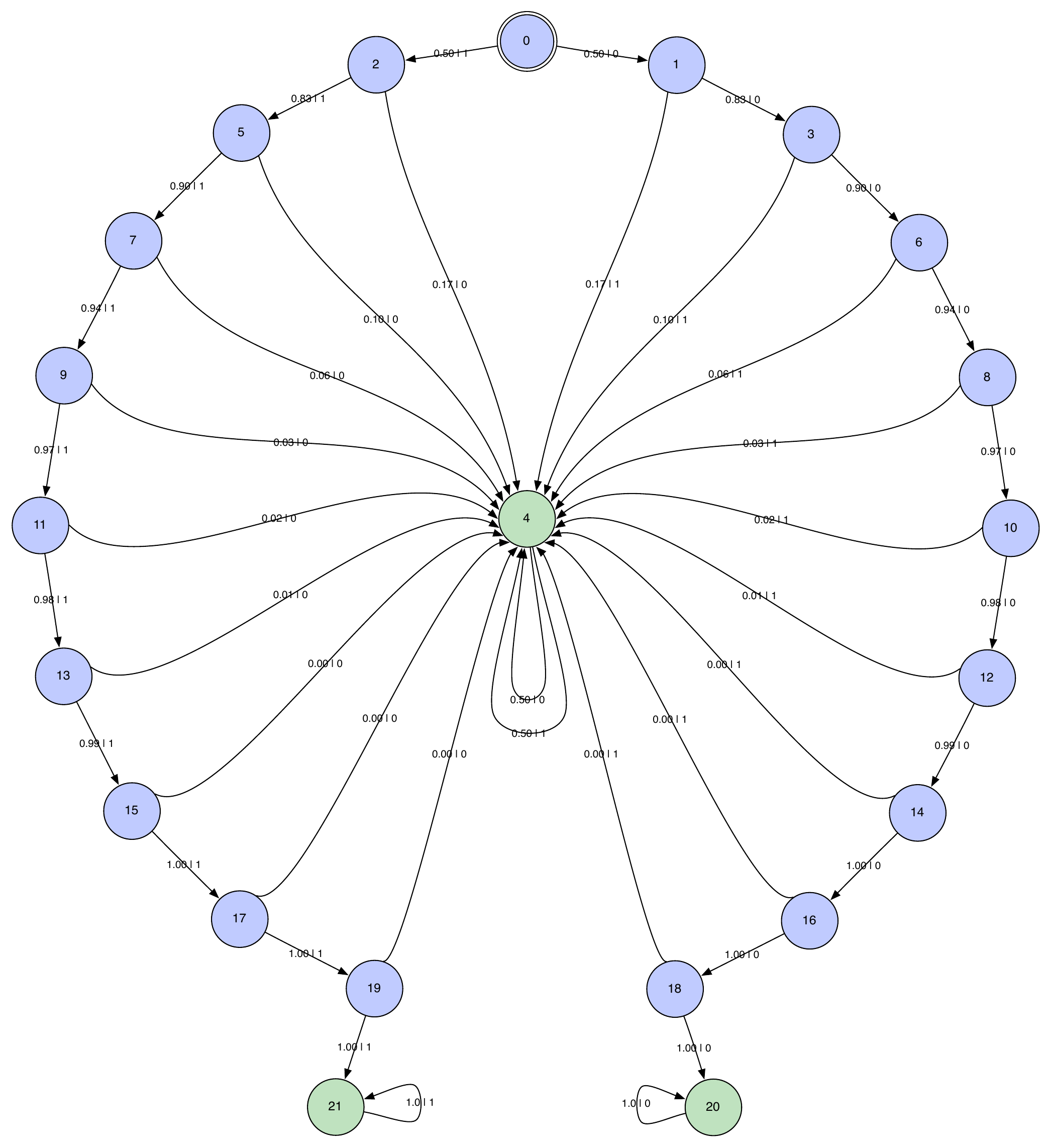}
\caption{The \eM\ HMSP presentation of two completely biased coins and one
	fair coin---all single-state \eMs: $M = \MSP_\pi \left(
	\text{All-}1\text{s} \times \text{All-}0\text{s} \times
	\Bernoulli{\tfrac{1}{2}} \right)$ with $\pi =
	\left(\tfrac{1}{3},\tfrac{1}{3},\tfrac{1}{3}\right)$.
  }
\label{fig:MOAPAll1sAll0sFC}
\end{figure}

To get a brief sense of the Purse Process consider an HMSP consisting of
three coins of unequal bias and compare this to the case of two coins of Sec.
\ref{sec:BCs}. Figure \ref{fig:MOAPAll1sAll0sFC} shows the HMSP for two
completely biased coins and one fair coin.

Its basic features have already encountered above. And, it suggests a notable
generalization that we now turn to.

\subsubsection{Mother of All Processes}
\label{sec:PU}

The generalization is to an HMSP consisting of a mixture of all processes.
Let's step through its construction.

First, recall that every stationary process has a unique \eM\ presentation. Thus, \eMs\ and stationary processes are in $1$-to-$1$ correspondence.
Second, an efficient algorithm exists to list all \eMs\ by the number of
the recurrent causal states.
Reference \cite{John10a} shows how to systematically enumerate
the \emph{\eM\ process library} $\mathcal{L}_k$ for $k$-state \eMs.
See Table I there for the list of binary-alphabet topological \eMs.
There are $1,117,768,214$ such $8$-state \eMs.

In the current construction consider only \emph{topological} \eMs\ for which any
branching transitions are taken with fair probability.
We refer to each process by it's \eM's enumeration number---we call
this the process' \emph{G\"{o}del} number.

Second, define the \emph{Process Urn} (PU) as containing the entire library of
\eMs. That is, we imagine a HMSP that is the result of reaching into the Urn,
selecting one \eM, and having it generate a full realization. The repeatedly
sampled PU is a HMSP---The Mother of All Processes. Certainly, one of the most
nonergodic processes one could work with.

\begin{Def}
The \emph{Mother of All Processes} is:
\begin{equation}
M = \MSP_\pi \left( \bigotimes_{M \in \mathcal{L}} M \right) ~,
\end{equation}
with $\pi$ being a chosen mixture distribution.
\end{Def}

To simplify, let's examine the HMSP whose components are all one-state
and all two-state \eMs. There are now $10$ components---three $1$-state
components and $7$ $2$-state components.

\begin{figure}
  \centering
  \includegraphics[width=\columnwidth]{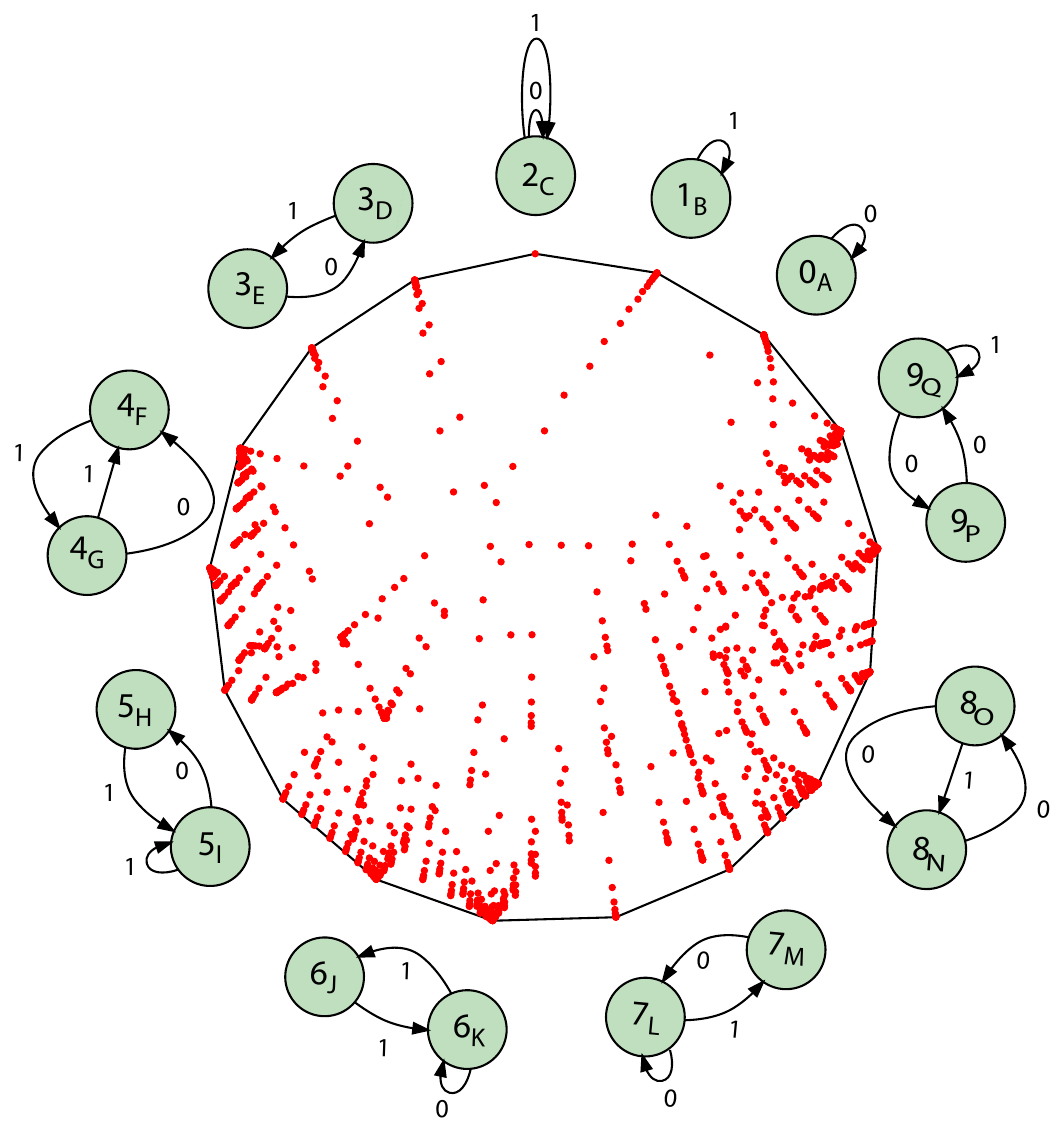}
\caption{The \eM\ presentation of the Mother of All $1$-State and
	$2$-State HMSP: $M = \MSP_\pi \left( \prod_{M \in
	\mathcal{L}_{\{1,2\}}} M \right)$ with $\pi = \left(\tfrac{1}{10},
	\ldots,\tfrac{1}{10}\right)$.  The start state is the $17$-vector at
	$\left( \tfrac{1}{10}, \tfrac{1}{10}, \tfrac{1}{10}, \tfrac{1}{20}, \ldots,
	\tfrac{1}{20} \right)$.  The \eMs\ for each ergodic component are placed
	around the periphery, dropping the transition probabilities which are
	either $1$ or $\tfrac{1}{2}$. Their states are aligned to their associated
	vertex in $\Delta^{17}$. The Fair Coin is the state at the top---the
	top-most vertex. The mixed states are approximated up to word lengths of
	$\ell = 28$.
  }
\label{fig:MOAP_All1s2s}
\end{figure}

There are $17$ recurrent causal states altogether across the ergodic
components. However, the many hundreds of mixed states are no longer usefully
presented in a state-transition diagram, as done up to this point. Instead, we
plot the mixed-states themselves as dots in the simplex $\Delta^{17}$.
This is shown in Fig. \ref{fig:MOAP_All1s2s}. This is a 2D projection in which
the recurrent states are the vertices of $\Delta^{17}$ and so appear on its
periphery. The start state, with uniform probability across the components and
not across the recurrent states, is not in the simplex center.

One notes the concentration of mixed states that move near to
$\Delta^{17}$'s vertices, indicating close approaches to synchronization.

There are a number of notable properties:
\begin{enumerate}
      \setlength{\topsep}{-4pt}
      \setlength{\itemsep}{-4pt}
      \setlength{\parsep}{-4pt}
\item The simplex vertices correspond to recurrent causal states.
\item There is an uncountably-infinite number of transient states. These fill
	out a complicated fractal measure within the $\Delta^{17}$.
\item All mixed states that are not on vertices are transient states.
\item While it is clear that $M$ is not exactly synchronizable \cite{Trav10a}
	as it contains infinite Markov-order components, is it asymptotically
	synchronizable \cite{Trav10b}? What about the synchronizability of
	approximations to it?
\end{enumerate}

%

There are a number of open questions, as well:
\begin{enumerate}
      \setlength{\topsep}{-4pt}
      \setlength{\itemsep}{-4pt}
      \setlength{\parsep}{-4pt}
\item What is $M$'s the statistical complexity dimension $d_\mu$
	\cite{Jurg20c}?
\item What is the shortest synchronizing word to go from the Fair Coin to each
	other ergodic component?
\item How do informational measures grow with word-length approximation?
\end{enumerate}

\section{Discussion}

In addition to their particular application, the ergodic decompositions give
important insight into basic questions about what structural complexity is and
how to measure it. A number of previous efforts that address these definitional
issues consider it a key property that complexity be additive over the
components of a system \cite{Benn86}. This is often motivated by a parallel
with Boltzmann entropy in thermodynamics. And, for that matter, additivity was
also posited as an axiom by Shannon for his measure of surprise \cite{Shan48a}.

However, the ergodic decompositions here show that the manner in which a
system's components relate to one another---specifically, the mixture
distribution---plays a central role in the process' organization and
contributes to quantitative measures of global complexity. The foregoing
offered a different, more structural view that goes beyond the ergodic
decompositions and statistical mixtures. Constructively, the transient state
structure is key to a multistationary process' global organization and what
observations can or cannot reveal.

The lessons here also suggest a skepticism in applying the ergodic
decompositions of Sec. \ref{sec:ErgDecomp}. One reason is that underlying them
is the assumption of an IID sampling of components, which is not generally
valid. Another is that they completely ignore how the internal structures of
the components interrelate with each other. And, as shown, this brings out
wholly new properties that are not part of any given component nor their sum
nor their IID mixture. Indeed, the mixture entropy does not capture this,
except in the most limited of cases.

Constructive responses to this will address the new kind of
hierarchical structure explicitly represented by the multistationary process'
\eM\ transient causal states and their complicated measure in $\Delta^{k}$.
Quantitatively, in contrast to the block entropy, entropy rate, and excess
entropy, we demonstrated that the transient information is sensitive to this
new kind of complexity in structural mixtures. It is this additional structure
that makes the organization of multistationary processes way more than the sum
of their parts. As a complementary metric, adapting the statistical complexity
dimension $d_\mu$ suggests itself \cite{Jurg20c}.

\section{Conclusions}

Let's close exploring several wider implications for thermodynamics, on the one
hand, and various attempts to introduce ``universal modeling'' schemes on the
other.

First, we started out highlighting the colloquialism, made familiar by the
social movements of the 1960s, that a system is more than the sum of its parts.
Presumably, the social reaction then reflected an increasing awareness of the
impact of technical systems humans were creating. The preceding development
explored in which senses this could be true for truly complex systems---ones
consisting many \emph{structured} components---more akin the social subsystems
than mere atoms. And, the various informational ergodic decompositions
bolstered the popular understanding.

However, in emphasizing structure and analyzing the concrete process class of
hidden multistationary processes it became abundantly clear---through all of
the examples presented---that composite or heterogeneous (to use Gibbs' word
\cite{Gibb75a}) systems are far more than the sum of their components.
Specifically, beyond a mere entropic, missing contribution from increased
disorder that arises from the random selection of components, composite systems
are markedly more complex. And, they are more structured according to the
interplay of the components' internal organization. It is that interplay
which drives the explosive complexity of multicomponent systems.

On this score, the history of composite systems is perhaps a bit confusing;
especially as they arose in the early foundations of thermodynamics. There is,
for example, Gibbs' seemingly contradictory statement, as quoted by Jaynes
\cite[p. 13]{Jayn92a}, that ``The whole is simpler than the sum of its parts''.
The ergodic decompositions seemed to say the opposite. However, there is not
really a confusion here. First, Gibbs was thinking of the correlations that
would emerge between system components when coupled together. Here, we
intentionally did not couple the components. Sequels address this. Second, at
root, the issue turns on an ambiguous vocabulary for describing randomness and
structure. Here, at least, by distinguishing between ``randomness'' in terms of
Shannon's notion of the flatness of a probability distribution and
``structure'' in terms of statistical complexity, we shed some light on these
important and still evolving issues.

Second, the HMSP construction procedure here gives a rather direct picture of
one kind of hierarchical organization in how a stochastic process can be built
from other processes. The constructive procedure uses the mixed state
presentation. And, this generates a new kind of hierarchy that emerges due to
the diverse combinatorial relationships between the components' internal
organizations. Other related hierarchies can be similarly construction; such as
when using generalized hidden Markov models \cite{Uppe97a} as ergodic
components.

Third and finally, modern statistical inference has been treated to a number of
formalizations of general learning that make minimal assumptions. Consider for
example:
\begin{enumerate}
      \setlength{\topsep}{-4pt}
      \setlength{\itemsep}{-4pt}
      \setlength{\parsep}{-4pt}
\item \emph{Universal Priors} \cite{Riss84a,Riss89a,Vita93a}: In the
	computation-theoretic approach to modeling and statistical inference there
	are attempts to define a most-general prior over model space. However,
	these raise very natural questions, What kind of process would generate
	such a prior?  Moreover, what kinds of difficulties are there in detecting
	processes drawn according to such a prior.
\item \emph{No Free Lunch Theorem} \cite{Wolp97a}: This framing makes a number
	of implicit assumptions about the measure on the Process Urn simplex. Does
	the theorem hold? Not when you consider structure.
\item \emph{Probably Almost Correct Learning} \cite{Vali84a}: This 
	``distribution-free'' approach is a bold attempt within machine learning to
	identify the computational nature of evolution and learning. However, is
	not this the same thing as assuming any process is possible? If so, then it
	is analogous to assuming the Mother of All Processes. That is, rather than
	being ``distribution-free'' the assumption underlying PAC learning is
	``distribution-full''.
\end{enumerate}

In light of these, The Mother of All Processes suggests a construction for such
assumption-free or minimal-assumption modeling. In this, one is sampling from
the space of all processes and exploits the \eM\ representation to be specific
about probability, on the one hand, and structure, on the other. The resulting
realization from this was that the preceding development was able to
demonstrate that transient-state structure made explicit the challenges in
detecting component processes.

\section*{Acknowledgments}

Many thanks to Chris Ellison for insightful discussions and coding and to the
Redwood Center for Theoretical Neuroscience, University of California at
Berkeley, for its hospitality during a sabbatical visit.

\bibliography{chaos}

\end{document}